\documentclass[aip,jcp,preprint,amsmath,amssymb]{revtex4-2}

\usepackage[utf8]{inputenc}
\usepackage[T1]{fontenc}
\usepackage{mathptmx}
\usepackage{etoolbox}
\usepackage{bm}
\usepackage{braket}
\usepackage{array}
\usepackage{booktabs}
\usepackage{graphicx}
\usepackage[caption=false]{subfig}
\usepackage{xcolor}
\usepackage{hyperref}
\usepackage{tabularx}
\hypersetup{hypertexnames=false,colorlinks=true,linkcolor=blue,citecolor=blue,urlcolor=blue}
\makeatletter
\def\@email#1#2{%
 \endgroup
 \patchcmd{\titleblock@produce}
  {\frontmatter@RRAPformat}
  {\frontmatter@RRAPformat{\produce@RRAP{*#1\href{mailto:#2}{#2}}}\frontmatter@RRAPformat}
  {}{}
}%
\makeatother

\begin{document}

\title{Counter-Electrode Dependence of the Working-Electrode Capacitance in Three-Electrode $TiO_2+rGO$ Cells: Classical Controls and Interpretation within a Quantum-Discord Framework}

\author{Kevin A. Gonzalez}
\author{Nicolás H. Toledo}
\author{David A. Miranda}
\email{dalemir@uis.edu.co}
\affiliation{Universidad Industrial de Santander, 680002 Bucaramanga, Santander, Colombia}

\date{\today}

\begin{abstract}
In three-electrode electrochemical impedance spectroscopy (EIS), the impedance/capacitance response of the working electrode (WE) is measured relative to the reference electrode (RE), whereas the current required by potentiostatic control flows between the WE and the counter electrode (CE). Under classical three-electrode operation, the measured WE--RE response should therefore be independent of the properties of the electrode used as the CE, provided that conventional electrochemical and instrumental artifacts are negligible. Here, we test this expectation in cells containing nanostructured electrodes $TiO_2+rGO$ and examine the results within a framework motivated by quantum discord to describe effective interelectrode correlations. We used dummy-cell measurements, a two-electrode bar--bar control, comparisons between Pt and bar CEs, a dummy circuit introduced into the CE branch, and CE ensemble comparisons to distinguish additive circuit behavior from CE-dependent behavior specific to the three-electrode configuration. In these experiments, the measured three-electrode capacitance response depends on CE, an effect that is not reproduced by the dummy-circuit controls examined. These findings identify a macroscopic CE-dependent capacitance response that departs from the tested classical expectation and is compatible with an effective interelectrode-correlation framework motivated by quantum discord, although the capacitance evidence alone does not constitute a direct measurement of quantum discord.
\end{abstract}

\maketitle

\section{Introduction}

Three-electrode electrochemical cells are designed to separate the control of the working-electrode (WE) potential from the passage of current through the counter electrode (CE) \cite{Wang2021Electrochemical}. In potentiostatic electrochemical impedance spectroscopy (EIS), a small sinusoidal perturbation superimposed on a DC bias is controlled between the WE and reference electrode (RE), while the current required to impose the prescribed WE--RE waveform flows between the WE and CE \cite{Bard2022,Snizhko2023Potentiostat,Wang2021Electrochemical}. Under normal operating conditions, the RE draws a negligible current, allowing the measured impedance to be interpreted as the WE--RE response and, under ideal conditions, assigned to the WE interface, provided that the CE supplies the required current without appreciable polarization and the potentiostat remains within its operating limits \cite{Vanysek2013The,Zhang2023}. This operational separation underpins the interpretation of EIS measurements at interfaces, in nanostructured materials, and in electrochemical devices \cite{Bard2022,Magar2021Electrochemical,Huang2016Graphical}. In a two-electrode configuration, on the contrary, the responses of both electrodes are intrinsically included in the measured impedance, whereas an ideal three-electrode configuration excludes the CE potential contribution from the voltage sensed between the WE and the RE \cite{Bard2022, Chakraborty2024Rectification}.

Importantly, assigning the measured response to the WE interface is an interpretation of the cell response, not an operating feature of the potentiostat that changes with the nature of the system under study \cite{Wang2021Electrochemical, Bard2022, Chakraborty2024Rectification}. At the macroscopic level, the instrument continues to sense the WE--RE potential and to drive the current between the WE and CE according to the same operating principle \cite{Wang2021Electrochemical,Bard2022,Lazanas2023Electrochemical}. Under ideal three-electrode conditions, this architecture supports the conventional interpretation that the measured WE--RE response can be treated independently of the physical state of CE \cite{Bard2022,Vanysek2013The,Wang2021Electrochemical,Zhang2023}. However, experimental evidence shows that CE can affect the measured WE response \cite{Bard2022,Chakraborty2024Rectification,Levi2014Impedance,Tatara2018The}. This dependence is conventionally attributed to identifiable limitations of CE, cell configuration, or potentiostatic control---including CE polarization, insufficient surface area or current-carrying capacity, counter-reaction kinetics, concentration polarization, mass-transport constraints, uncompensated solution resistance, RE placement, and control-loop limitations---that modify the electrochemical environment in WE or reduce the accuracy with which the intended WE--RE potential is imposed and recorded \cite{Bard2022,Vanysek2013The,Lazanas2023Electrochemical,Snizhko2023Potentiostat}. Moreover, the coupling of electron transfer, electrical-double-layer charging, and ion transport can produce distinct frequency-dependent impedance signatures \cite{Li2022,Yamamoto2024}. Although these mechanisms do not require the potentiostat to measure the CE impedance directly, they show that the processes occurring at WE and CE do not need to be entirely independent \cite{Wang2021Electrochemical,Lazanas2023Electrochemical,Tatara2018The}. Consequently, if the WE response depends systematically on the physical state of the CE while the potentiostat continues to sense only the WE--RE potential, this dependence suggests an effective correlation between the two electrodes, although it does not by itself establish the nature of that correlation \cite{Levi2014Impedance,Chakraborty2024Rectification}. Here, we propose that a CE-dependent response that remains after the tested instrumental artifacts and conventional electrochemical contributions have been assessed can be interpreted in terms of effective quantum correlations between the WE and CE, through which the physical state of the CE is reflected in the WE response without altering the macroscopic operating principle of the potentiostat \cite{RevModPhys.84.1655,Wang2019Nonequilibrium}. The central question is therefore whether such a systematic CE-dependent response persists under the experimental controls implemented in this study.

We examine this question within a quantum-statistical framework, using information-theoretic measures to characterize effective correlations between the electrodes \cite{Ollivier2001_Quantum,RevModPhys.84.1655,Luo2008Quantum}. In the effective description adopted here, the relevant degrees of freedom between the electrodes are associated with WE and CE, allowing the electrochemical cell to be modeled as a bipartite quantum system \cite{Wang2019Nonequilibrium,Miao2023Entanglement}. For such a system, quantum mutual information quantifies the total correlations and, with respect to a specified local measurement, can be decomposed into a classically accessible contribution and a residual nonclassical contribution identified with quantum discord \cite{Wang2019Nonequilibrium,Ollivier2001_Quantum}. In its general formulation, quantum discord can remain nonzero even for separable mixed states, in which entanglement is absent \cite{Ollivier2001_Quantum,RevModPhys.84.1655,Adesso2016Measures}. As an initial formal construction of the framework proposed here, the effective WE--CE state is modeled as a coherent superposition of states, yielding a pure bipartite state \cite{Ollivier2001_Quantum}. Within this pure-state approximation, every nonproduct state has nonzero discord, and the derivation presented below establishes the relationships among quantum mutual information, classically accessible correlations, and quantum discord. This deliberate first approximation provides the mathematical starting point of the framework and can subsequently be extended to mixed states, thereby encompassing separable states with nonzero discord while preserving the present pure-state construction as its foundation \cite{Ollivier2001_Quantum,RevModPhys.84.1655}.

The experimental observations reported by Miranda, Pinzon, and Bueno \cite{Miranda2024a} provide the phenomenological basis for this analysis. Whereas that work focused primarily on the experimental behavior and its quantum-motivated interpretation, the present study develops a more formal physical and mathematical description aimed specifically at clarifying the physical origin of this counter-electrode dependence. To this end, it distinguishes the classical correlations produced by additive two-electrode measurements from the CE independence expected in an ideal three-electrode measurement and formulates an effective bipartite WE--CE state for describing departures from that expectation. This progression connects the macroscopic control experiments to the information-theoretic interpretation developed in the following.

To explore the proposed quantum-statistical framework experimentally, we use nanostructured electrodes $TiO_2+rGO$ as model electrochemical interfaces. The nanostructured $TiO_2$ is particularly suitable for this purpose because its interfacial capacitance, surface chemistry, and charge-transport properties are strongly dependent on morphology and surface functionalization \cite{Miranda2021Nanostructured,Shen2018Titanium,Berger2012The}. Incorporation of reduced graphene oxide (rGO) provides an electronically active layer that can modify both differential capacitance and carrier dynamics \cite{Jadoon2024Hybrid,Tayebi2019Reduced}. Consequently, the measured interfacial capacitance may reflect contributions from electrical-double-layer charging, geometric and chemical capacitance, and quantum-capacitance effects associated with the electronic density of states \cite{Luryi1988QuantumCapacitance,Bisquert2015,Wu2022Understanding,Gadipelli2023Understanding,DiPasquale2023,Li2023}. More broadly, studies of graphene--electrolyte interfaces have shown that electrolyte structure, interfacial polarization, and electronic properties jointly influence the resulting capacitance, while recent descriptions of AC transport illustrate how ionic and electronic degrees of freedom can become coupled through electrochemical interfaces \cite{DiPasquale2023,Li2023,Coquinot2026}.

In the bipartite approximation adopted here, electrolyte-mediated effects at each interface are retained in the effective WE and CE states, whereas the bulk electrolyte---treated as a passive ionic conductor---and the ideally non-current-carrying RE are not introduced as separate quantum subsystems. Thus, the explicit quantum description is restricted to the effective WE and CE interfacial degrees of freedom; the detailed assumptions underlying this treatment are specified in the theoretical methods.

The experimental objective is therefore to determine whether the three-electrode EIS response of a $TiO_2+rGO$ WE remains independent of the electrode used as the CE and, if not, whether the observed dependence is accounted for by the classical contributions examined. We first establish the instrumental three-electrode reference using passive dummy circuits and the additive reference using two-electrode $TiO_2+rGO$ measurements. We then compare the three-electrode responses obtained with the Pt and $TiO_2+rGO$ CEs and introduce a passive dummy element into the CE branch as an additional control. Throughout this analysis, the impedance and derived complex-capacitance spectra are treated as macroscopic observables used to test the adequacy of the classical independent-impedance description, not as direct measurements of a quantum state or quantum discord. A CE-dependent response that persists across the tested controls is therefore interpreted as evidence compatible with the proposed effective quantum-discord framework rather than as a direct experimental determination of discord.

\section{Theoretical Details and Experimental Methods}

This section details the theoretical and experimental methods used to study how the response of the electrochemical cell depends on the counter electrode. We begin by defining the impedance measured in two- and three-electrode setups and by introducing the proposed effective description of the cell within a statistical-mechanical framework. Next, we apply classical and quantum information theory to construct the corresponding correlation measures and to establish baseline expectations for the effective WE--CE system. Finally, we outline the fabrication of the $TiO_2+rGO$ electrodes, the EIS conditions and potentiostat settings, the experimental measurement configurations, and the data transformations used to analyze the impedance and capacitance spectra.

\subsection{Electrochemical Impedance Measurements}

For a potentiostatic EIS measurement, the applied perturbation can be written as $v(t)=V_0\sin(\omega t)+V_1$, and the measured current response as $i(t)=I_0\sin(\omega t+\phi)+I_1$. The measured complex impedance is given by $Z_m(\omega)=\frac{V_0}{I_0}e^{-j\phi}$.

The measured impedance was transformed into complex capacitance according to $C^*(\omega)=1/[j\omega Z_m(\omega)]=C'(\omega)-jC''(\omega)$, using the convention $Z_m(\omega)=Z'(\omega)-jZ''(\omega)$ \cite{ITAGAKI2007}. Consequently, the complex-capacitance Nyquist representations plot $C''$ against $C'$, while the Bode representations show $C'$ and $C''$ as functions of frequency. The same transformation and sign convention was applied to each configuration.

In a two-electrode configuration, the CE and RE electrodes are short-circuited, and the measured impedance therefore reflects the combined contribution of both electrodes. Because the electrodes used in this study were manufactured in the form of small bars, the term bar--bar is used solely as a convenient descriptor of the experimental arrangement. Under these conditions, two impedances can be associated with the system: one corresponding to the working electrode bar, including interfacial effects, $Z_{\mathrm{WE}}(\omega)$, and the other corresponding to the counter electrode, $Z_{\mathrm{CE}}(\omega)$. The total impedance of the electrochemical cell in a two-electrode configuration is thus described by Eq.~(\ref{eq:two_electrode}).

\begin{equation}
\label{eq:two_electrode}
Z_m^{(2p)}(\omega)=Z_{WE}(\omega)+Z_{CE}(\omega).
\end{equation}

Consequently, in two-electrode measurements, a nonzero classical dependence of $Z_m^{(2p)}$ on $Z_{CE}$ is expected because the impedance of the CE contributes directly to the measured observable. By contrast, in a three-electrode potentiostatic configuration, the measured response is assigned to the WE--RE branch and is given by Eq.~(\ref{eq:three_electrode}).

\begin{equation}
\label{eq:three_electrode}
    Z_m^{(3p)}(\omega)=Z_{WE}(\omega).
\end{equation}

In the three-electrode configuration, CE is not measured directly by the potentiostat. The measured impedance corresponds to the response of the WE with respect to the RE, while the CE serves solely to complete the current path with the WE. Therefore, in this work, the quantities indicated as $Z_{WE}$ and $Z_{CE}$ refer to the impedance responses obtained from independent characterization measurements of each bar. In both the two-electrode and three-electrode bar--bar experiments, each bar was first characterized separately under the same individual three-electrode condition, using Pt as CE and Ag/AgCl as RE. After this independent characterization, one bar was connected as WE and the other bar was connected either to the shorted RE/CE leads in the two-electrode bar--bar measurement or as CE in the three-electrode bar--bar measurement. Thus, when CE material is discussed in the bar--bar three-electrode experiment, $Z_{CE}$ does not represent an impedance measured simultaneously during that experiment. Instead, it represents the independently measured impedance response of the same bar that was later connected to the CE.

\subsection{Classical-Information Framework}

For classical random variables $X$ and $Y$, the correlations can be quantified by mutual information $I(X;Y)$, given by Eq.~(\ref{eq:classical_mi}), where $p(x,y)$ is the joint probability distribution, $p(x)$ and $p(y)$ are the corresponding marginal distributions, and $p(x|y)=p(x,y)/p(y)$ for $p(y)>0$ is the conditional probability. It should be noted that while $X$ and $Y$ are classical random variables corresponding to measured signals (e.g., electrode readouts), this probabilistic formulation remains framework-agnostic, with $p(x,y)$ arising from either a classical model or quantum state collapse. Mutual information is non-negative and vanishes only when the joint distribution is factorized into the product of its marginals \cite{Lesne2014Shannon}.

\begin{equation}
\label{eq:classical_mi}
I(X;Y)=\sum\limits_{x,y}p(x,y)\ln\!\left[\frac{p(x,y)}{p(x)p(y)}\right].
\end{equation}

An alternative way to quantify correlations is via the classically accessible correlation $J(X|Y)$, defined in Eq.~(\ref{eq:classical_acc_corr}), where the Shannon entropy of the random variable $X$ is $H(X)=-\sum\limits_x p(x)\ln p(x)$ and the conditional Shannon entropy is $H(X|Y)=-\sum\limits_{x,y}p(x,y)\ln p(x|y)$. Within a strictly classical probabilistic framework, the mutual information $I(X;Y)$ coincides with the classically accessible correlation $J(X|Y)$. Consequently, the notion of quantum discord arises precisely because these two quantities can differ in the presence of genuinely quantum correlations. Throughout this manuscript, the semicolon notation $I(X;Y)$ is used to denote mutual information between classical random variables, including experimentally measured impedance time series. In contrast, the colon notation $I(A:B)_\rho$ is reserved for quantum mutual information between subsystems of an effective bipartite quantum state $\rho$.

\begin{equation}
\label{eq:classical_acc_corr}
J(X|Y)=H(X)-H(X|Y).
\end{equation}

\subsubsection{Classical Two-Electrode Expectation}
\label{Classical 2-Electrode Expectation}

The two-electrode configuration provides the classical reference case in which the counter-electrode contribution is intentionally included in the measured observable. In probabilistic terms, the relevant random variables are the measured two-electrode impedance, $Z_m^{(2p)}$, and the independently characterized counter-electrode impedance, $Z_{CE}$. Even if $Z_{WE}$ and $Z_{CE}$ are statistically independent before the two-electrode constraint is imposed, the measured variable $Z_m^{(2p)}=Z_{WE}+Z_{CE}$ contains the CE contribution as one of its components. Therefore, the joint distribution between $Z_m^{(2p)}$ and $Z_{CE}$ should not be factored.

For the two-electrode configuration, mutual information can be written as
\begin{equation}
\label{eq:mi_two_electrode}
I\!\left(Z_m^{(2p)};Z_{CE}\right)
=
\sum\limits_{z_m^{(2p)},z_{CE}}
p\!\left(z_m^{(2p)},z_{CE}\right)
\ln
\left[
\frac{
p\!\left(z_m^{(2p)},z_{CE}\right)
}{
p\!\left(z_m^{(2p)}\right)
p\!\left(z_{CE}\right)
}
\right].
\end{equation}
Under the additive model, the measured variable is constrained by $z_m^{(2p)}=z_{WE}+z_{CE}$. For discrete variables, this constraint can be represented as
\begin{equation}
\label{eq:additive_joint_constraint}
p\!\left(z_m^{(2p)},z_{CE}\right)
=
\sum\limits_{z_{WE}}
p(z_{WE})p\!\left(z_{CE}\right)
\delta\!\left(z_m^{(2p)}-z_{WE}-z_{CE}\right).
\end{equation}

Consequently, the classical expectation is $I\!\left(Z_m^{(2p)};Z_{CE}\right)>0$ when variations in $Z_{CE}$ are experimentally distinguishable and are not fully masked by WE variability or measurement uncertainty. Because this is a classical probability model, the classically accessible correlation equals the mutual information,
\begin{equation}
\label{eq:two_electrode_classical_correlation}
J\!\left(Z_m^{(2p)}|Z_{CE}\right)
=
I\!\left(Z_m^{(2p)};Z_{CE}\right).
\end{equation}
The full derivation is provided in Sect. S1 of the Supplementary Material, Eqs. (S1)--(S7). Thus, a nonzero correlation in Configuration 2 is not anomalous; it is the expected classical result of measuring an observable that includes the counter-electrode impedance.

\subsubsection{Classical Three-Electrode Expectation}

The three-electrode configuration defines the classical reference case in which the CE closes the current path but its impedance is not part of the measured WE--RE observable. Under ideal three-electrode operation, the measured impedance is assigned to the WE--RE branch, Eq.~(\ref{eq:three_electrode}).

Let repeated EIS measurements be described by random variables representing the measured three-electrode response, $Z_m^{(3p)}$, and the independently characterized impedance state of the electrode later used as CE, $Z_{CE}$. If the response follows the classical three-electrode expectation, then
\begin{equation}
\label{eq:three_electrode_factorization}
p\!\left(z_m^{(3p)},z_{CE}\right)
=
p_m^{(3p)}\!\left(z_m^{(3p)}\right)
p_{CE}\!\left(z_{CE}\right),
\end{equation}
and therefore
\begin{equation}
\label{eq:three_electrode_zero_correlation}
J\!\left(Z_m^{(3p)}|Z_{CE}\right)
=
I\!\left(Z_m^{(3p)};Z_{CE}\right)
=0.
\end{equation}
The complete information-theoretic derivation is given in Sect. S2 of the Supplementary Material, Eqs. (S8)--(S14). The practical classical reference expectation is therefore

\begin{equation}
\label{eq:delta_ce_reference}
\Delta Z_{CE}(\omega)=Z_m^{(3p)}(\omega|CE=TiO_2+rGO)-Z_m^{(3p)}(\omega|CE=Pt)=0,
\end{equation}

within experimental uncertainty and within the range over which classical CE artifacts have been excluded.

\subsection{Quantum-Information Framework}

To formulate the hypothesis of nonclassical interelectrode correlations, we introduce an effective bipartite state $\hat{\rho}_{WE,CE}$ associated with interfacial degrees of freedom in the WE and in the electrode used as the CE. The CE state is not measured directly by the potentiostat, which is configured to measure the WE response relative to the RE. Instead, the bipartite state provides an effective quantum-information description of the interfacial electronic degrees of freedom whose macroscopic response is probed through the EIS measurement.

Within this framework, three limiting situations can be distinguished. First, if $\hat{\rho}_{WE,CE}=\hat{\rho}_{WE}\otimes\hat{\rho}_{CE}$, the effective WE-CE state is a product state and does not contain correlations. Second, if the state is separable but not factorable, it contains no entanglement, although quantum correlations, such as discord, may still be present. Finally, if the state is nonseparable, then entanglement is present, yet additional types of quantum correlations may also be present \cite{Ollivier2001_Quantum}\;. Therefore, separability only guarantees the absence of entanglement; it does not, by itself, rule out nonclassical correlations.

The total correlations in the effective state are quantified by the quantum mutual information $I(WE:CE)_\rho=S(\hat{\rho}_{WE})+S(\hat{\rho}_{CE})-S(\hat{\rho}_{WE,CE})$, where $S(\hat{\rho})=-\mathrm{Tr}(\hat{\rho}\ln\hat{\rho})$ is the von Neumann entropy. The classically accessible part of these correlations is denoted $J(CE:WE)_\rho$. The discord associated with these measurements, $\mathcal{D}(CE:WE)_\rho$, is given by Eq.~(\ref{eq:discord}).

\begin{equation}
\label{eq:discord}
\mathcal{D}(CE:WE)_\rho=I(WE:CE)_\rho-J(CE:WE)_\rho.
\end{equation}

A nonzero value of $\mathcal{D}(CE:WE)_\rho$ indicates that effective interelectrode correlations cannot be fully represented as classically accessible correlations. In the present work, this formulation is used to interpret whether counter-electrode-dependent three-electrode impedance responses are compatible with nonclassical interelectrode correlations. It does not imply that $Z_m$, $Z_{CE}$, or characteristic times are themselves quantum states. These quantities are macroscopic observables that are used to test whether the classical three-electrode expectation is sufficient.

Here, we model the electrochemical cell as a coherent superposition of quantum states. The electronic structures of both the WE and CE are described in terms of electrons and holes exchanged at their respective electrode--electrolyte interfaces. The WE and CE provide the path for current flow, whereas the RE establishes the measurement reference and ideally carries no current, i.e., the measurement is performed between WE and RE. Therefore, the RE need not be included as an explicit subsystem in the quantum description. Because the WE and CE are distinct components of the electrochemical cell, corresponding to electrodes with appropriately modified surfaces, each can be characterized by its own Hamiltonian.

Within the frequency range of interest, we assume that the electrolyte can be treated as a passive ionic conductor that (i) does not mediate effective correlations between the WE and CE, (ii) introduces no appreciable ionic fluctuations into the interelectrode dynamics under consideration, (iii) contributes only a classical dissipative background without significantly affecting the coherence of the relevant interfacial degrees of freedom, and (iv) exhibits no appreciable intrinsic frequency-dependent response. Under these assumptions, the explicit quantum description can be restricted to the WE and CE subsystems.

This approximation is particularly relevant when the supporting electrolyte is selected to provide sufficient ionic conductivity while minimizing specific adsorption, parasitic faradaic reactions, concentration polarization, and mass-transport limitations at the electrode interfaces. It may be considered reasonable within an experimentally identified intermediate-frequency window in which the bulk electrolyte contribution can be approximated by a frequency-independent solution resistance, while low-frequency diffusion and polarization effects and high-frequency dielectric or instrumental effects remain negligible.

In this framework, let $\mathcal{H}_{WE}$ and $\mathcal{H}_{CE}$ denote the Hilbert spaces associated with the WE and CE, respectively. We choose orthonormal bases
\begin{equation}
\left\{ \ket{WE_i} \right\} \subset \mathcal{H}_{WE},
\qquad
\left\{ \ket{CE_j} \right\} \subset \mathcal{H}_{CE},
\end{equation}
and the subsystem states are written as

\begin{equation}
\ket{WE}=\sum_i w_i \ket{WE_i},
\qquad
\ket{CE}=\sum_j c_j \ket{CE_j}.
\end{equation}

Then, we consider the bipartite pure state

\begin{equation}
\label{eq:ansatz}
\ket{\Psi} = \sqrt{p}\,\ket{S} + \sqrt{1-p}\,\ket{\mathcal{C}}, \qquad 0 \le p \le 1,
\end{equation}
where $\ket{S} = \ket{WE}\ket{CE}$ is a fully separable contribution and $\ket{\mathcal{C}} = \sum_{i,j} a_{ij}\ket{WE_i}\ket{CE_j}$ is a general correlated contribution, both normalized, with $\mathrm{Re}\Big(\sum_{i,j}w_i^*c_j^*a_{ij}\Big)=0$ ensuring $\braket{\Psi|\Psi}=1$ for every $p$ (see Appendix~\ref{app:pure_state_derivation}). The weight $p$ interpolates between the fully factorized limit ($p=1$) and the limit in which the state carries only the correlated component ($p=0$).

Because the density operator of the composite system, $\hat{\rho}_{WE,CE}=\ket{\Psi}\bra{\Psi}$, is pure for every value of $p$ (equivalently, $\hat{\rho}_{WE,CE}^{2}=\hat{\rho}_{WE,CE}$ \cite{blum2012density}), this ansatz can realize only the first and third regimes discussed above: for a pure state, separability is the same as being a product state, so the intermediate regime of states that are separable but not factorizable cannot occur. For $p\in[0,1]$, the state is exactly factored in $p=1$ for arbitrary $a_{ij}$, and it is factored in $p=0$ if and only if $a_{ij}$ has Schmidt rank one \cite{nielsen_chuang_2010}. For a generic $a_{ij}$ with Schmidt rank greater than one, the state remains entangled for all $0<p<1$, although special, non-generic choices of $a_{ij}$ might yield isolated values of $p$ where the state becomes factorizable. The purity condition also enforces $S(\hat{\rho}_{WE,CE})=0$ identically, and the Schmidt decomposition of $\ket{\Psi}$ ensures $S(\hat{\rho}_{WE})=S(\hat{\rho}_{CE})$, so that we obtain $I(WE:CE)_\rho$ from Eq.~(\ref{eq:mutual-info}), as derived in Appendix~\ref{app:pure_state_derivation}.

\begin{equation}
\label{eq:mutual-info}
I(WE:CE)_\rho = 2\,S(\hat{\rho}_{CE}).
\end{equation}

For the classically accessible correlation, we consider projective measurements $\{\Pi_k^{WE}\}$ on the part of the effective bipartite system associated with the WE, matching the potentiostat's access to the cell through the WE--RE response. A direct calculation in Appendix~\ref{app:pure_state_derivation} shows that, for any measurement basis and any outcome $k$, the conditional CE state $\hat{\rho}_{CE|k}$ remains pure, a structural consequence of the global purity of $\hat{\rho}_{WE,CE}$, independent of $p$ or $a_{ij}$. The conditional entropy therefore vanishes identically, every measurement basis is equally optimal, and
\begin{equation}
\label{eq:classical-corr}
J(CE:WE)_\rho = S(\hat{\rho}_{CE}).
\end{equation}
Combining Eqs.~(\ref{eq:mutual-info}) and~(\ref{eq:classical-corr}) via Eq.~(\ref{eq:discord}) gives
\begin{equation}
\label{eq:discord-value}
\mathcal{D}(CE:WE)_\rho = S(\hat{\rho}_{CE}) \ge 0.
\end{equation}

At $p=1$, $S(\hat{\rho}_{CE})=0$ and $I(WE:CE)_\rho = J(CE:WE)_\rho = \mathcal{D}(CE:WE)_\rho = 0$, recovering the classical expectation that a WE-only measurement carries no information about the CE. For any $0<p<1$ with $a_{ij}$ of Schmidt rank greater than one, $S(\hat{\rho}_{CE})>0$ and all three quantities are strictly positive; within this ansatz, nonzero discord is therefore the generic prediction whenever interelectrode correlations are present rather than a fine-tuned exception. The limiting cases and the role of the Schmidt rank are summarized in the Appendix~\ref{app:pure_state_derivation}.

\subsection{Fabrication of \texorpdfstring{$TiO_2+rGO$}{TiO2+rGO} Electrodes}

Cylindrical $TiO_2$ bars were modified on one base with graphene oxide followed by electrochemical reduction. The base assigned for the coating was wet-sanded using 5000-grit paper for 25 min and then sequentially polished using suspensions of 0.3, 0.1 and 0.05 $\mu$m for 25, 10, and 5 min, respectively. After polishing, the samples were immersed in isopropyl alcohol and sonicated for 25 min to remove surface residues.

\begin{figure}[t]
    \centering
    \includegraphics[width=0.68\textwidth]{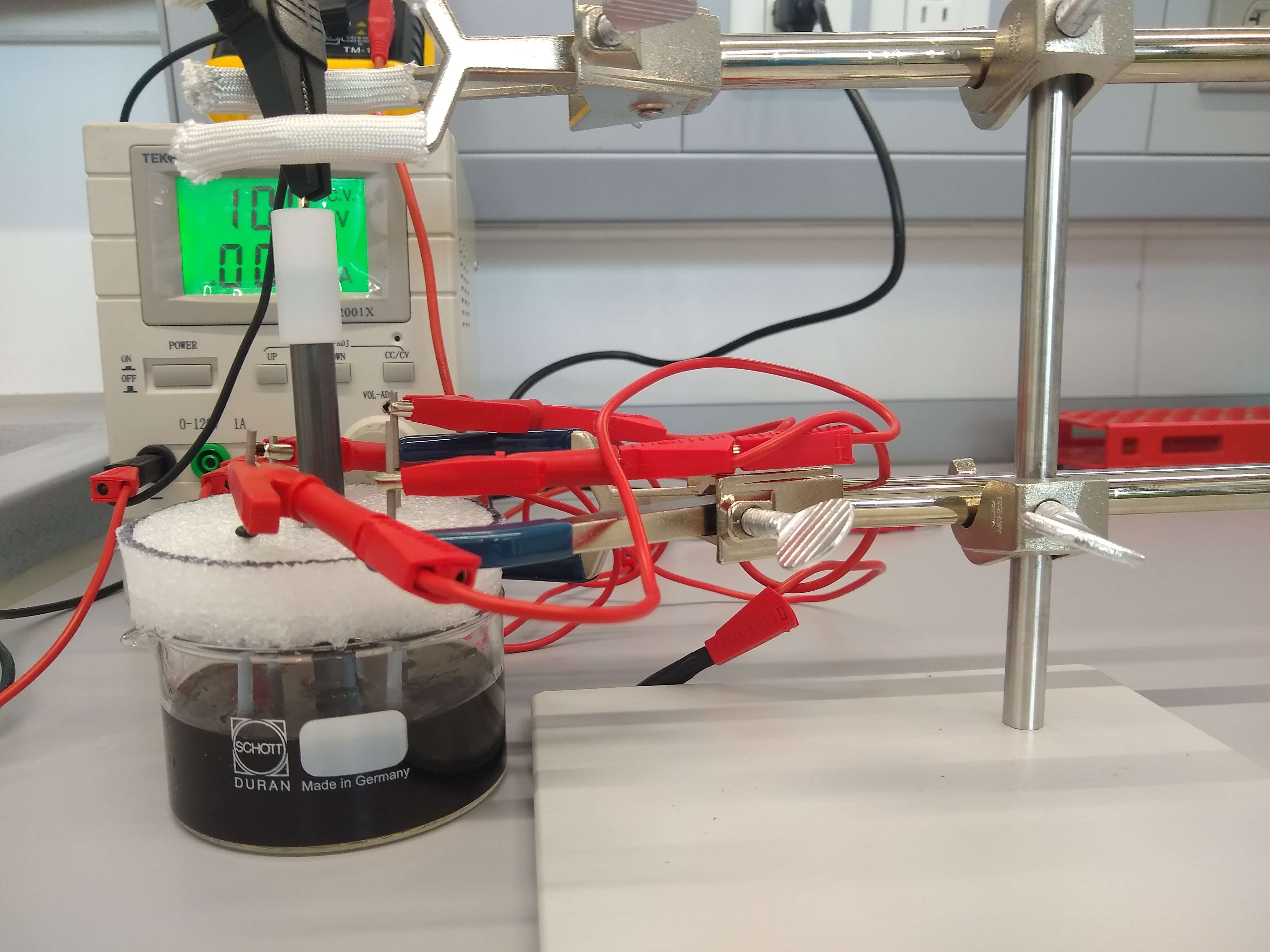}
    \caption{Experimental arrangement used during graphene oxide deposition on cylindrical $TiO_2$ bars. The preparation sequence included surface polishing, sonication, graphene oxide deposition, rinsing/drying, and electrochemical reduction to obtain $TiO_2+rGO$ electrodes.}
    \label{fig_electrodeposition_setup}
\end{figure}

Graphene oxide suspensions were prepared at 0.05 g mL$^{-1}$ using 5 g of graphene oxide in 100 mL of a 90:10 v/v ethanol/type-I water mixture. The graphene oxide powder was obtained from Graphene Investments Science and Innovation, with a reported purity of 99\% and product code GO-890321. The ethanol-rich mixture was used to reduce water electrolysis during deposition. The suspension was homogenized by sonication for 20 min in an ultrasonic bath. Graphene oxide was deposited onto an exposed face of Teflon-coated titanium bars in a two-electrode configuration using the TiO$_2$ bar as WE and graphite as CE. A 10 V DC potential was applied for 3 min while monitoring the current; see Fig.~\ref{fig_electrodeposition_setup}. The electrode was then removed, rinsed and dried. The formation of a brown-yellow film on the polished TiO$_2$ surface was used as visual evidence of the deposition of graphene oxide.

The coated bars were then electrochemically reduced in a three-electrode cell containing 3 mL of phosphate-buffered saline (PBS). The graphene-oxide-coated $TiO_2$ bar was connected as WE, Pt was used as CE, and an Ag/AgCl electrode with 3 M KCl internal solution was used as RE. Ten cyclic-voltammetry scans were applied between $-1.2$ and $0.4$ V to reduce the deposited graphene oxide and increase the electronic conductivity of the coating. The final electrodes are denoted $TiO_2+rGO$. A total of 16 bars were fabricated following the procedure described above, and all 16 were subjected to individual EIS characterization using Configuration 1 (see Sect.~\ref{sec:configurations}). Six bars were excluded from subsequent comparisons because their characterization measurements yielded responses that could not be analyzed reliably across configurations, including spectra with unexpected or anomalous features. The remaining ten bars were grouped into five pairs by matching bars whose individual Configuration 1 measurements, obtained with Pt as CE, showed comparable impedance spectra. The pairing was therefore based on the Pt-CE characterization rather than on the outcome of the subsequent Configuration 3 comparison.

The main text presents measurements from one of these pairs, for which the dummy-element control was also performed. The other four pairs were measured in Configurations 1 and 3 without the dummy element and are presented in the Supplementary Material. These five pairs constitute the complete set of comparisons obtained from the ten bars retained for analysis. The absolute capacitance magnitudes varied among the bars, likely due to small differences in rGO deposition, coating uniformity, exposed area after polishing, and local conductivity.

\subsection{EIS Conditions and Potentiostat Control}

All EIS measurements were performed in potentiostatic mode using a 0.01 V$_{\mathrm{RMS}}$ AC perturbation, with the open-circuit potential used as the DC reference. The frequency range was 1 MHz to 0.1 Hz with 80 measurement points. The experiments were carried out at approximately 19 $^\circ$C. Electrochemical measurements were performed in 3 mL of phosphate-buffered saline (PBS, pH 7.4), which was used as a supporting electrolyte to study the electrode/electrolyte interface. A Metrohm-Autolab PGSTAT-204 potentiostat/galvanostat equipped with an FRA-32M module and controlled by NOVA 2.1 software was used for all measurements. The same Ag/AgCl reference electrode was employed throughout the experiments, and the electrochemical cell was placed inside a Faraday cage to reduce electromagnetic noise.

The potentiostat operation was verified using a dummy-cell arrangement. The physical sequence was CE lead, CE-branch test network, RE lead, dummy cell, and WE lead. The dummy-cell terminal used corresponded to an approximately 1 k$\Omega$ resistance in parallel with an approximately 1 $\mu$F capacitance. Eight configurations were measured: direct dummy-cell connection, four CE-branch resistors (1, 3, 10 and 47 k$\Omega$), and three CE-branch RC networks using a 1 $\mu$F capacitor in series with selected resistors. In the ideal potentiostatic configuration, the measured spectrum should remain determined by the dummy cell between RE and WE rather than by the elements inserted before the RE in the CE branch.

\begin{figure*}[t]
\centering
\subfloat[Configuration 1: individual three-electrode characterization.\label{config1}]{%
\includegraphics[width=0.4\textwidth]{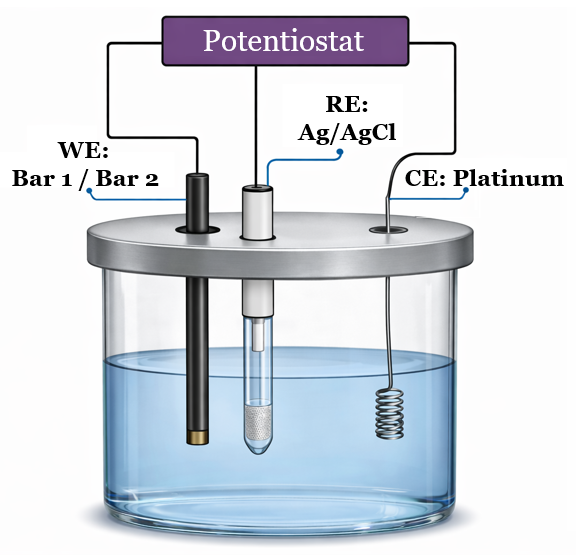}}
\hfill
\subfloat[Configuration 2: two-electrode bar--bar measurement.\label{config2}]{%
\includegraphics[width=0.4\textwidth]{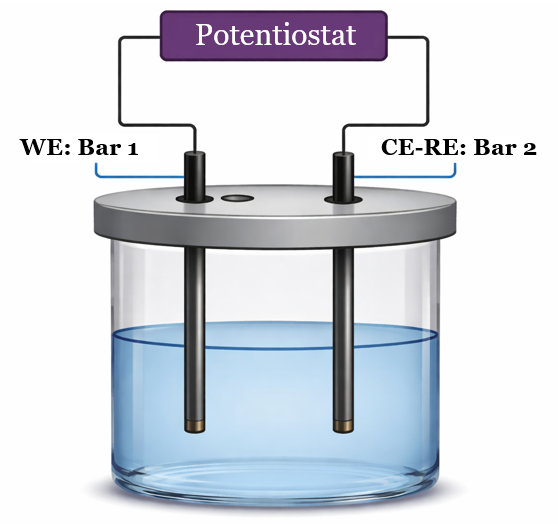}}

\subfloat[Configuration 3: three-electrode bar--bar measurement.\label{config3}]{%
    \includegraphics[width=0.4\textwidth]{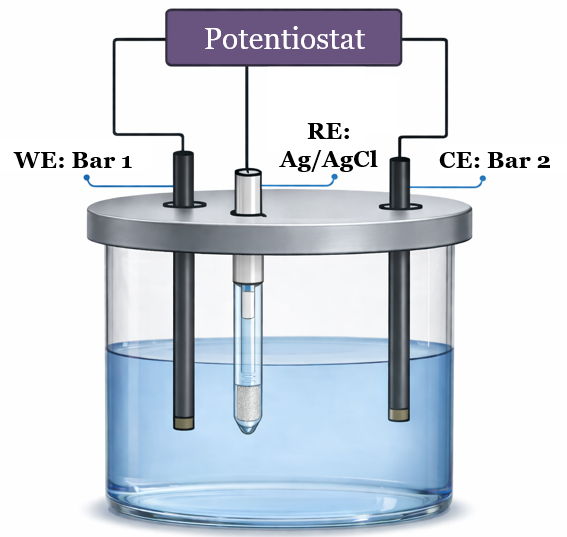}}
\caption{Diagrams of the electrochemical cells representing the three configurations used for the EIS measurements.}
\label{tresconfiguraciones}

\end{figure*}

\subsection{Measurement Configurations}
\label{sec:configurations}

\begin{figure*}[t]
\centering
\subfloat[]{%
\label{fig_Configuration2withDummy}
\includegraphics[width=0.35\textwidth]{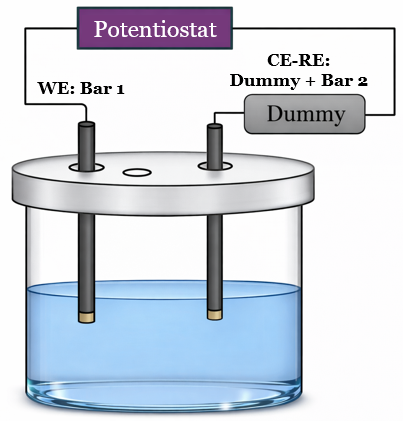}}
\hfill
\subfloat[]{%
\label{fig_Configuration3withDummy}
\includegraphics[width=0.35\textwidth]{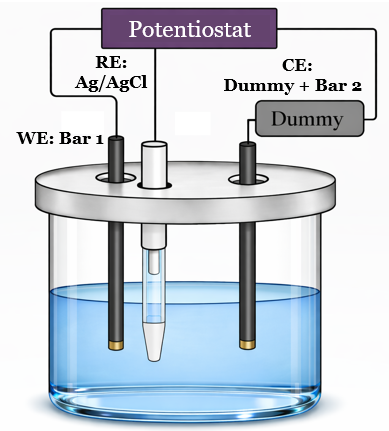}}
\caption{Modified experimental arrangements used for the dummy-cell control. (a) Configuration 2 with a passive dummy cell connected within the measured two-electrode series path. This arrangement provides the positive classical reference in which the dummy-cell impedance should appear in the measured response. (b) Configuration 3 with the same dummy cell introduced into the CE branch while the Ag/AgCl electrode remains as the RE. This arrangement tests whether the passive classical contribution enters the WE--RE response when it is confined to the CE current path.}
\label{fig_dummy_configurations}
\end{figure*}

Three different measurement configurations were used to distinguish the classical additive behavior from the CE-dependent response of three-electrodes, as summarized in Fig.~\ref{tresconfiguraciones}. In Configuration 1, see  Fig.~\ref{config1}, each $TiO_2+rGO$ bar was individually characterized as the WE using Pt as the CE and Ag/AgCl as the RE. These measurements yield the individual electrode responses that serve as the basis for subsequent comparisons. In Configuration 2, Fig.~\ref{config2}, a $TiO_2+rGO$ bar was connected as the WE, while a second bar was connected to the shorted RE/CE leads, so the resulting two-electrode response is expected to reflect the additive series contribution of both bars. Finally, in Configuration 3, Fig.~\ref{config3}, one bar was connected as the WE and the other as the CE, with Ag/AgCl retained as the RE. This three-electrode configuration evaluates whether the measured WE--RE response depends on the independently characterized state of the bar employed as the CE.

To determine whether the CE-dependent response could be reproduced by a system exhibiting purely classical circuit behavior, a passive dummy cell was introduced into Configurations 2 and 3, as shown in Fig.~\ref{fig_dummy_configurations}. In modified Configuration 2, Fig.~\ref{fig_Configuration2withDummy}, the dummy cell lies within the measured two-electrode series path and should therefore produce its ordinary classical impedance signature, providing a positive classical control. Configuration 3 was then modified by introducing the same dummy cell into the CE branch of the three-electrode bar--bar arrangement, Fig.~\ref{fig_Configuration3withDummy}, in which the CE-dependent response compatible with effective nonclassical interelectrode correlations is observed. Under the proposed interpretation, the passive dummy cell provides only a classical circuit contribution and does not supply the correlated electrochemical interfacial degrees of freedom represented by the CE subsystem. Its impedance signature should therefore follow the conventional three-electrode expectation and should not appear in the measured WE--RE response. The persistence of the bar-specific CE-dependent response together with the disappearance of the dummy-cell signature would indicate that the observed behavior is not a generic consequence of adding a classical impedance to the CE branch.

\section{Results and Discussion}

The results are organized as a progression from instrumental and additive classical reference cases to the three-electrode CE-dependence test. Passive resistive and RC networks first establish whether a contribution placed in the CE branch enters the measured WE--RE response. Configuration 2 then tests the expected series addition using two independently characterized $TiO_2+rGO$ bars, whereas the comparison of Configurations 1 and 3 determines whether replacing the Pt CE with a $TiO_2+rGO$ bar changes the response of the same WE. The dummy-in-CE experiment finally tests whether a passive impedance in the CE branch can reproduce the observed change.

Throughout this progression, the measured impedance $Z_m(\omega)$ and the derived complex capacitance $C^*(\omega)$ are treated as macroscopic observables for evaluating classical reference cases. A CE-dependent response is considered within the effective quantum-information framework only after it has been assessed against the passive-circuit controls; the spectra themselves are not treated as quantum states or direct measurements of quantum discord.

\subsection{Instrumental Baseline: Dummy Cell}

Before performing the electrochemical measurements, we validated the three-electrode measurement geometry using the passive-circuit arrangement described in the Methods, with a fixed dummy cell in the WE--RE branch and eight direct, resistive, or RC configurations in the CE branch.

Under ideal potentiostatic operation, the measured response should therefore be determined by the fixed dummy cell and remain independent of the networks placed between the CE and the RE. This control tests the instrumental premise required to interpret the subsequent electrode-dependent measurements.

\begin{figure}[t]
\centering
\includegraphics[width=\textwidth]{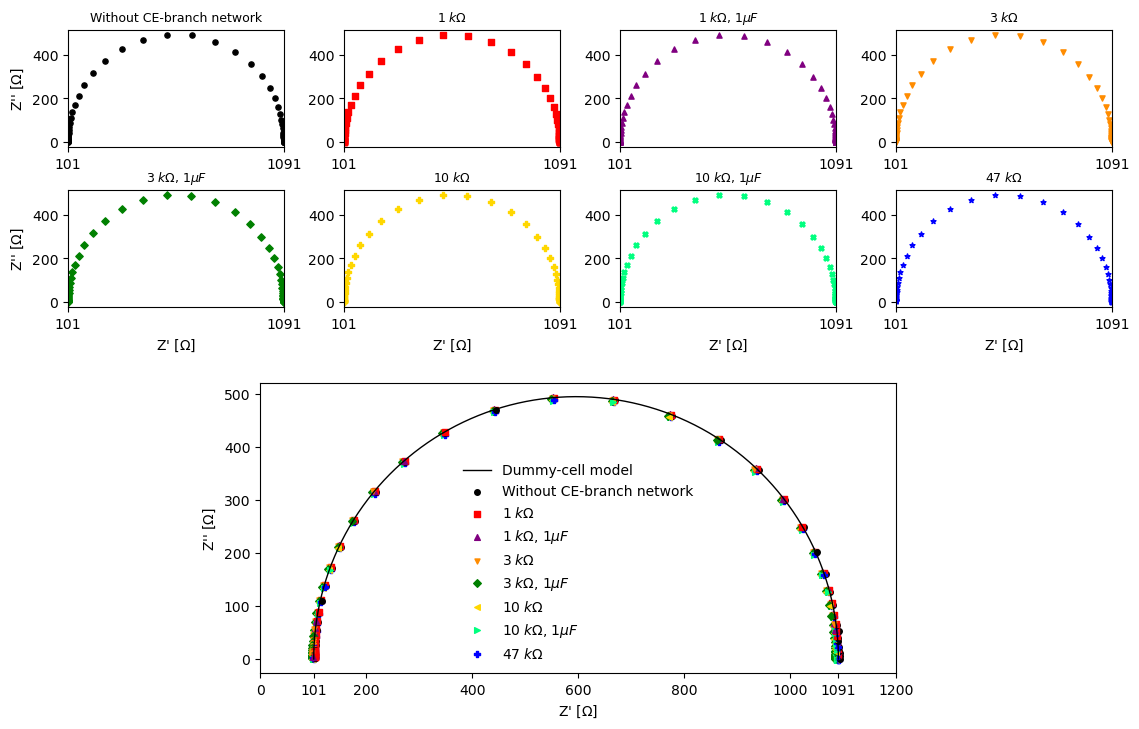}
\caption{Instrumental validation using passive networks in the CE path. The Nyquist spectra of the fixed dummy cell in the WE--RE branch are shown for the direct connection, four resistive networks, and three RC networks placed before the RE in the CE branch. All eight responses coincide within the graphical resolution; a small offset between corresponding points was introduced solely for visualization.}
\label{fig_potentiostat_validation}
\end{figure}

The Nyquist curves for the dummy-cell were visually superposed for all eight CE-branch configurations (Fig.~\ref{fig_potentiostat_validation}). The main semicircle extended from approximately 100 $\Omega$ to 1100 $\Omega$, giving $R_{\infty}\approx100\,\Omega$, $R_s\approx1100\,\Omega$, and $\Delta R\approx1000\,\Omega$, consistent with the nominal resistance of the dummy. The response was described as a Debye-like relaxation, $Z_{\mathrm{dummy}}(\omega)=R_{\infty}+\Delta R/(1+j\omega\tau)$, whose vertex satisfies $\omega\tau=1$. Because the CE-branch network did not produce visually distinguishable changes in either $\tau$ or $\Delta R$, the observed spectra follow the single semicircle predicted by this model.

This purely classical control establishes the instrumental precondition for the later CE-dependent comparisons: with the RE correctly positioned, passive networks placed exclusively in the CE current path do not appear in the measured WE--RE response. It does not invoke the effective quantum-information framework because these networks do not provide the interfacial electronic degrees of freedom represented by the WE and CE subsystems. A subsequent dependence on the electroactive electrode used as CE can therefore be distinguished from the simple entry of a passive CE-branch contribution into the measurement chain.

\begin{figure*}[t]
\centering
\includegraphics[width=\textwidth]{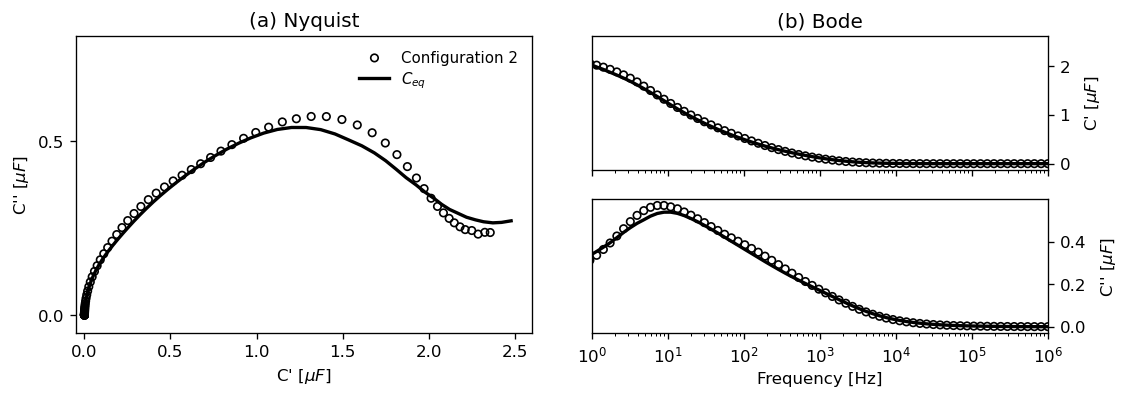}
\caption{The measured two-electrode bar-–bar response is compared with the equivalent response calculated from two independently characterized bars. (a) Complex-capacitance Nyquist plot ($C''$ vs. $C'$). (b) Bode plots of $C'$ (top) and $C''$ (bottom) vs. frequency. The qualitative match supports the classical additive interpretation of Configuration 2.}
\label{fig_two_electrode_ceq}
\end{figure*}

\subsection{Two-Electrode Additivity: Bar--Bar Configuration}

The first electrochemical comparison follows directly from the classical correlation framework. In the two-electrode configuration, a dependence on the CE is expected because $Z_{CE}$ enters the measured observable itself. The bar--bar measurement therefore serves as a classical control rather than as a test of non-classical correlations. After the two bars are independently characterized in Configuration 1, they are measured together in Configuration 2. Their two-electrode impedance should then follow Eq.~(\ref{eq:two_electrode}). Expressed in terms of the complex capacitance of two elements connected in series, the calculated equivalent response is

\begin{equation}
\label{eq:ceq}
C_{eq}(\omega)
=
\frac{C_1(\omega)C_2(\omega)}
{C_1(\omega)+C_2(\omega)}.
\end{equation}

The measured Configuration 2 response agrees qualitatively with the equivalent response calculated from the independently characterized bars (Fig.~\ref{fig_two_electrode_ceq}). Because no quantitative spectral-distance analysis was performed, the comparison is interpreted qualitatively rather than assigning a statistical level of agreement. Nevertheless, the available comparison shows that the experimental procedure recovers the expected classical series behavior when the CE contribution is intentionally included in the measured two-electrode observable. It therefore establishes the positive classical-correlation reference against which the three-electrode expectation is subsequently evaluated.

Within the information-theoretical framework, this qualitative agreement is precisely what the classical model predicts. Because $Z_{CE}$ enters $Z_m^{(2p)}$ directly through Eq.~(\ref{eq:two_electrode}), $I\left(Z_m^{(2p)};Z_{CE}\right)>0$ follows from the functional relationship between the two variables, independent of any additional correlation between the WE and CE interfaces. This remains true even though both electrodes are $TiO_2+rGO$ bars: Configuration 2 cannot determine whether their interfacial degrees of freedom contain additional classical or nonclassical correlations, but only establishes that the additive model is sufficient when the CE contribution is included in the measured quantity. The discriminating comparison is therefore Configuration 3, for which the ideal classical three-electrode model does not predict a dependence on $Z_{CE}$.

\subsection{Three-Electrode Counter-Electrode Dependence: Bar--Pt versus Bar--Bar}

\begin{figure*}[t]
\centering
\includegraphics[width=\textwidth]{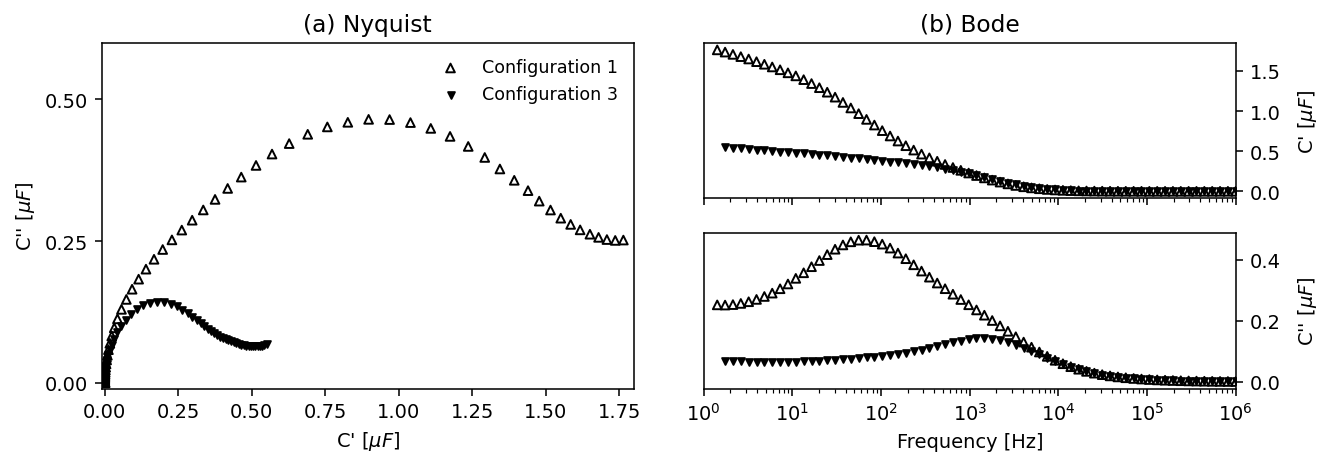}
\caption{Three-electrode counter-electrode-dependent response for the $TiO_2+rGO$ bar set used in the subsequent dummy-cell control (Sect.~\ref{sec:dummy_element_control}). The same $TiO_2+rGO$ WE exhibits different WE--RE responses when Pt and a second $TiO_2+rGO$ bar are used as CE. (a) Complex-capacitance Nyquist representation, $C''$ versus $C'$. (b) Bode representation showing $C'$ in the upper panel and $C''$ in the lower panel as functions of frequency. The difference between the two configurations constitutes the primary deviation addressed in the manuscript. The displayed frequency range extends from 1 MHz down to approximately 1 Hz.}
\label{fig_pt_vs_barbar_3e}
\end{figure*}

The central three-electrode comparison tests whether the measured WE--RE response remains independent of the electrode used as CE. According to the classical three-electrode expectation, the same $TiO_2+rGO$ WE measured with Pt as CE and with a second $TiO_2+rGO$ bar as CE should produce the same response within experimental uncertainty. Instead, replacing the Pt CE with a $TiO_2+rGO$ bar modifies the measured three-electrode response (Fig.~\ref{fig_pt_vs_barbar_3e}). This difference is summarized by $\Delta Z_{CE}(\omega)$ in Eq.(~\ref{eq:delta_ce_reference}). The four other valid bar pairs were compared under Configurations 1 and 3 without the dummy element, and their complete pairwise results are provided in Sect.~S4 and Figs.~S1--S4 of the Supplementary Material. In conjunction with the main-text pair, these measurements comprise the complete set of bar–bar comparisons derived from the ten bars retained for analysis.

Unlike the two-electrode configuration, in which $Z_{CE}$ enters the measured observable directly, the ideal three-electrode reference assigns $Z_m^{(3p)}=Z_{WE}$ to the WE--RE branch while the RE carries a negligible current. A nonzero value of $\Delta Z_{CE}(\omega)$ therefore represents a departure from this classical reference rather than the trivial consequence of measuring an observable that already contains $Z_{CE}$.

Pt is a metal with a large and nearly continuous density of electronic states, and its behavior as a CE is consistent with the classical current-source picture assumed in the conventional three-electrode operating principle. By contrast, the $TiO_2+rGO$ bar used as CE in Configuration 3 is the same nanostructured material used as WE. As discussed in the Introduction, its interfacial capacitance can contain a contribution associated with the electronic density of states through quantum capacitance. Quantum capacitance is a property of each individual interface and does not, by itself, imply interelectrode discord. However, it indicates that the electronic structure of the $TiO_2+rGO$ interface can contribute directly to its interfacial capacitance and therefore provides a physical basis for retaining these electronic degrees of freedom in effective description. Representing the WE and CE as an effective bipartite system in which interelectrode correlations may be considered is, therefore, a physically motivated hypothesis rather than an arbitrary construction.

Within the effective quantum-information framework, this CE-dependent response is consistent with the correlated regime of the proposed ansatz. The classical three-electrode expectation is associated with the product-state limit $p=1$, for which $\mathcal{D}(CE:WE)_\rho=0$ and a WE-only measurement carries no information about the CE. By contrast, dependence on the independently characterized CE state is compatible with $\mathcal{D}(CE:WE)_\rho=S(\hat{\rho}_{CE})>0$, for which the effective WE and CE interfacial states contain correlations that cannot be fully represented by a classical WE-only description. Under this interpretation, the observed response is a macroscopic signature compatible with nonzero effective discord, rather than a direct measurement of it.

The asymmetry between Pt and $TiO_2+rGO$ can be connected more specifically to their electronic densities of states. Quantum capacitance scales with the density of states at the Fermi level, $C_Q\propto D(E_F)$, and combines in series with the double-layer capacitance according to $1/C_{\mathrm{total}}=1/C_{dl}+1/C_Q$. Consequently, a sufficiently large $D(E_F)$ renders the quantum-capacitance contribution effectively irrelevant to the measured response, regardless of the finer electronic structure beneath it.

Within the present framework, this relation offers a plausible physical interpretation of the classical limit. Even if a coupling channel between the WE and CE interfacial states were present when Pt was used as CE, its large and nearly continuous density of states could leave no distinguishable signature in the measured observable. This behavior is compatible with $\mathcal{D}(CE:WE)_\rho\approx0$ and with an effective state close to the product-state limit. The CE $TiO_2+rGO$, whose more limited carrier density preserves the structure in the density of states over the relevant energy range, is not subject to the same suppression and is therefore consistent with a regime in which $\mathcal{D}(CE:WE)_\rho$ does not need to vanish. This connection is offered as a plausible physical interpretation of the observed asymmetry, not as a derivation of $p$ from the microscopic electronic structure, which lies beyond the scope of the present framework.

The following control examines whether this CE-dependent signature can instead be reproduced by a simpler classical contribution placed in the CE branch.

\subsection{Dummy-Element Control in the Counter-Electrode Branch}
\label{sec:dummy_element_control}

The dummy-element control was evaluated in two successive steps. First, the CE-dependent response of the set of bars used in this experiment was established by comparing Configuration 1, in which Pt was used as the CE, with Configuration 3, in which a second $TiO_2+rGO$ bar was used as the CE. As shown in Fig.~\ref{fig_pt_vs_barbar_3e}, the two three-electrode responses differ, providing the bar-specific CE-dependent reference against which the subsequent dummy-cell control is evaluated.

In the second step, a dummy cell expected to exhibit classical behavior was incorporated into modified Configurations~2 and~3, as shown in Fig.~\ref{fig_dummy_configurations}, to determine how an additional passive classical contribution affects the measured response in each configuration. In modified Configuration 2, the dummy cell lies within the measured two-electrode series path and should therefore contribute its capacitance signature directly to the measured response. This configuration serves as the positive control for detecting the characteristic dummy-cell relaxation. In modified Configuration 3, the same dummy cell is placed in series with the same $TiO_2+rGO$ CE but remains confined to the CE branch, while the potentiostat continues to measure the WE--RE response. Under the classical three-electrode expectation, the characteristic dummy-cell relaxation should not be resolved in the WE--RE response. Moreover, if the bar-specific response observed in Configuration 3 is associated with effective nonclassical interelectrode correlations, the addition of a passive classical element in the CE branch is not expected to appreciably modify that response. The comparison therefore tests whether a passive classical contribution in the CE branch can reproduce or appreciably perturb the bar-specific response compatible with effective nonclassical interelectrode correlations.

\begin{figure*}[t]
\centering
\includegraphics[width=\textwidth]{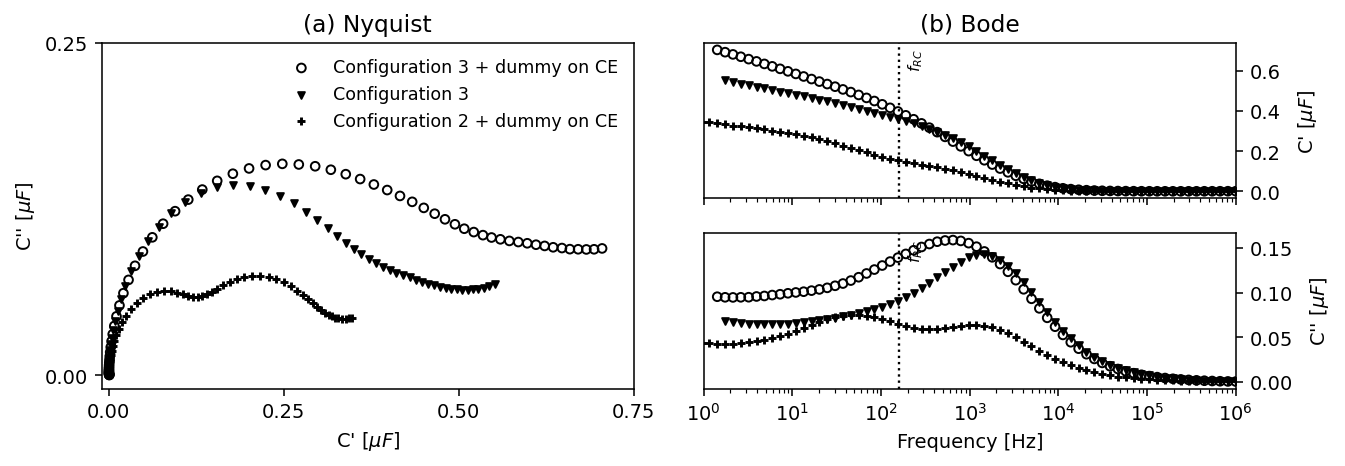}
\caption{Dummy-element control in the counter-electrode branch. Curves are identified by the symbols in the legend. (a) Complex-capacitance Nyquist plot ($C''$ versus $C'$). Configuration 2, with the dummy cell in the measured two-electrode path, exhibits two relaxation lobes, including an additional low-frequency lobe at larger $C'$. (b) Bode plots of $C'$ (top) and $C''$ (bottom). The corresponding low-frequency contribution in $C''$ is not resolved in Configuration 3, where the dummy cell is confined to the CE branch and the response remains dominated by the higher-frequency relaxation observed without the dummy cell. The dotted line marks the nominal dummy-circuit characteristic frequency, $f_{\mathrm{dummy}}=159.15~\mathrm{Hz}$, calculated from the component values. The two Configuration 3 spectra differ slightly; residual differences near the dominant relaxation are not assigned to a specific origin. Frequency range: $1~\mathrm{MHz}$ to approximately $1~\mathrm{Hz}$.}
\label{fig_dummy_ce_control}
\end{figure*}

Configuration 2 provides the expected positive-control response: when the dummy cell lies within the measured two-electrode series path, two resolved relaxation lobes are observed (Fig.~\ref{fig_dummy_ce_control}), unlike the other measurements reported in the paper and the Supplementary Material. The dummy circuit consists of a $100~\Omega$ resistor in series with a parallel $1~\mathrm{k}\Omega$--$1~\mu\mathrm{F}$ network, for which $\tau_{\mathrm{dummy}}=R_pC=1~\mathrm{ms}$ and $f_{\mathrm{dummy}}=(2\pi R_pC)^{-1}=159.15~\mathrm{Hz}$. Because the dummy impedance and the electrochemical impedance enter the Configuration 2 observable in series, their separated characteristic times generate distinguishable relaxation features \cite{Hahn2019}. The transformation from impedance to complex capacitance is nonlinear; consequently, $f_{\mathrm{dummy}}$ is shown as the nominal frequency calculated from the component values rather than as a fitted position of the maximum in $C''$ \cite{ITAGAKI2007}.

When the same dummy cell is transferred to the CE branch of Configuration 3, the additional low-frequency relaxation observed in Configuration 2 is no longer resolved. Both Configuration 3 spectra remain dominated by a single higher-frequency relaxation near $1~\mathrm{kHz}$. Nevertheless, modest differences in the amplitude, width, and position of the dominant relaxation remain between Configuration 3 with and without the dummy cell. The available measurements do not establish the origin of these residual differences. Accordingly, the conclusion of this control is restricted to the non-resolution of the characteristic dummy-cell relaxation in the WE--RE response and does not imply an exactly zero effect of inserting the dummy cell.

If the CE-dependent response established in the first step resulted from indiscriminate mixing of CE-branch contributions into the WE--RE measurement by the potentiostatic control loop, placing the dummy cell in that branch should introduce the distinct low-frequency relaxation detected in the positive two-electrode control. Its absence as a resolved feature in Configuration 3 argues against such indiscriminate mixing and indicates that the persistent CE-dependent response is associated with the electroactive $TiO_2+rGO$ interface rather than with the mere presence of a passive impedance in the CE branch.

Within the effective quantum-information framework, a passive dummy cell does not provide the electrochemical interfacial degrees of freedom represented by the CE subsystem. The fact that its characteristic relaxation is not resolved in the WE--RE response is consistent with the classical three-electrode expectation, whereas persistence of the $TiO_2+rGO$-specific response is compatible with $\mathcal{D}(CE:WE)_\rho>0$. This control therefore supports the proposed quantum-motivated interpretation by showing that the observed CE dependence is not reproduced by the tested passive classical contribution. Nevertheless, it does not constitute a direct determination of quantum discord.

\section{Conclusions}

This work tested whether the WE--RE capacitance response in a three-electrode cell remains independent of the CE after accounting for conventional instrumental and additive circuits. Dummy-cell experiments showed that passive resistive and RC elements confined to the CE branch did not introduce their characteristic relaxation into the measured WE--RE response under the tested conditions. In a two-electrode setup, pairs of $TiO_2+rGO$ electrodes qualitatively reproduced the series combination expected from independently characterized electrodes. These measurements define the classical instrumental and additive reference cases needed to interpret the three-electrode experiments.

However, replacing the Pt CE with a second electrode $TiO_2+rGO$ altered the response of the three-electrodes for the same WE. Adding a passive dummy element to the CE branch did not reproduce this change: the dummy contribution appeared as an additional resolved relaxation when placed directly in the two-electrode path, but was not resolved in the corresponding three-electrode WE--RE response. Although modest differences remained between the Configuration 3 spectra with and without the dummy cell, they did not reproduce its characteristic low-frequency contribution. The observed CE dependence therefore cannot be explained by simple series addition of independently characterized impedances or by indiscriminate inclusion of a passive CE-branch impedance. Instead, it departs from the ideal independent-impedance description and shows that the measured WE response depends on the specific electroactive CE interface.

Within the effective quantum-information framework proposed here, this macroscopic CE-dependent response is consistent with a nonproduct WE--CE state for which the proposed pure-state ansatz predicts nonzero quantum discord. This consistency does not constitute a direct measurement of quantum discord: the impedance and capacitance spectra are macroscopic observables, the model parameter $p$ is not extracted from the present data, and the controls do not rule out all conventional electrochemical explanations. The results instead provide a controlled phenomenological basis and a testable hypothesis for interpreting CE-dependent impedance via effective nonclassical interelectrode correlations. More conclusive evaluation will require quantitative uncertainty analysis across the available CE ensemble, additional tests of conventional electrochemical contributions, and a microscopic link between the measured response and the effective quantum state.

\section*{Supplementary Material}

The Supplementary Material provides complete derivations of the classical two-electrode additive reference and the ideal classical three-electrode independence reference, along with extended algebraic details underpinning the normalization, reduced-state, and conditional-measurement calculations summarized in Appendix~\ref{app:pure_state_derivation}. It further reports the full set of Configuration 1--Configuration 3 comparisons for the remaining four valid bar pairs (Figs.~S1--S4), thereby complementing the representative pair presented in the main text. No dummy element was incorporated in these supplementary measurements.

\begin{acknowledgments}
The authors acknowledge the \textit{Universidad Industrial de Santander} for financial support provided to Kevin A. Gonzalez through a master’s scholarship. The authors also thank Dr. Edgar Fabián Pinzón Nieto for critically reviewing an earlier version of the manuscript and for providing constructive feedback that improved the discussion of reproducibility, control experiments, and the interpretation of the dummy-element results.

During manuscript preparation, the authors used ChatGPT and Codex (OpenAI) and Writefull for Overleaf (Digital Science). ChatGPT and Writefull supported English-language editing. ChatGPT was also used exclusively to refine the graphical appearance of author-drawn schematic illustrations, without altering their scientific content or interpretation. All data plots were generated from the experimental data using Python, without generative-AI assistance. Codex was used to organize and document the accompanying GitHub repository. All AI-assisted outputs were critically reviewed and verified by the authors.
\end{acknowledgments}

\section*{Author Declarations}

\subsection*{Conflict of Interest}

The authors have no conflicts to disclose.

\subsection*{Author Contributions}

All authors contributed to Conceptualization, Formal analysis, Investigation, Methodology, Software, and Writing--review \& editing. Nicolas H. Toledo and Kevin A. Gonzalez contributed to Data curation, Visualization, and Writing--original draft. David A. Miranda contributed to Funding acquisition, Project administration, Resources, Supervision, and Validation.

\section*{Data Availability}

The raw EIS data underlying all data-driven figures reported in this work are publicly available, including measurements of the ten $TiO_2+rGO$ bars retained for analysis and the passive dummy-cell controls, together with the Python code and notebook used to reproduce the corresponding figures. The dataset and the code are archived in Zenodo (version 1.0.0, DOI: \href{https://doi.org/10.5281/zenodo.22178264}{10.5281/zenodo.22178264}) \cite{GTM2026a} and are also available through the associated GitHub repository at \url{https://github.com/davidalejandromiranda/electrochemical-cell-nonclassical-correlations}. Raw characterization files are not available for the six excluded bars.

\appendix

\section{Formal Derivation of the Effective Pure-State WE--CE Correlation Relations}
\label{app:pure_state_derivation}

Let $\mathcal{H}_{WE}$ and $\mathcal{H}_{CE}$ denote the Hilbert spaces associated with the effective interfacial degrees of freedom of the WE and CE, respectively, and let $\{\ket{WE_i}\}$ and $\{\ket{CE_j}\}$ be orthonormal bases. The subsystem states are
\begin{equation}
\ket{WE}=\sum_i w_i\ket{WE_i},
\qquad
\ket{CE}=\sum_j c_j\ket{CE_j},
\label{eq:app_subsystem_states}
\end{equation}
with $\sum_i|w_i|^2=\sum_j|c_j|^2=1$. We define the normalized product and correlated contributions as
\begin{equation}
\ket{S}=\ket{WE}\ket{CE},
\qquad
\ket{\mathcal{C}}=\sum_{i,j}a_{ij}\ket{WE_i}\ket{CE_j},
\qquad
\sum_{i,j}|a_{ij}|^2=1.
\label{eq:app_components}
\end{equation}
The effective pure-state ansatz is
\begin{equation}
\ket{\Psi}=\sqrt{p}\,\ket{S}+\sqrt{1-p}\,\ket{\mathcal{C}},
\qquad 0\leq p\leq 1.
\label{eq:app_ansatz}
\end{equation}
Writing $z=\braket{S|\mathcal{C}}=\sum_{i,j}w_i^*c_j^*a_{ij}$ gives
\begin{equation}
\braket{\Psi|\Psi}
=1+2\sqrt{p(1-p)}\,\operatorname{Re}(z).
\label{eq:app_normalization}
\end{equation}
Thus, the condition $\operatorname{Re}(z)=0$ ensures normalization for every $p$. Expanded component-level algebra supporting the normalization, partial-trace, and conditional-measurement calculations in this Appendix is provided in Sect. S3 of the Supplementary Material.

Defining
\begin{equation}
B_{mn}=\sqrt{p}\,w_m c_n+\sqrt{1-p}\,a_{mn},
\label{eq:app_Bmn}
\end{equation}
the state and density operator can be written as
\begin{equation}
\ket{\Psi}=\sum_{m,n}B_{mn}\ket{WE_m}\ket{CE_n},
\qquad
\hat{\rho}_{WE,CE}=\ket{\Psi}\bra{\Psi}.
\label{eq:app_density_operator}
\end{equation}
Taking the corresponding partial traces yields
\begin{align}
\hat{\rho}_{WE}
&=\sum_{m,m'}\left(\sum_n B_{mn}B_{m'n}^*\right)
\ket{WE_m}\bra{WE_{m'}},
\label{eq:app_rho_we}\\
\hat{\rho}_{CE}
&=\sum_{n,n'}\left(\sum_m B_{mn}B_{mn'}^*\right)
\ket{CE_n}\bra{CE_{n'}}.
\label{eq:app_rho_ce}
\end{align}
Because $\hat{\rho}_{WE,CE}$ is constructed from a normalized state vector,
\begin{equation}
\hat{\rho}_{WE,CE}^2=\hat{\rho}_{WE,CE},
\qquad
S(\hat{\rho}_{WE,CE})=0.
\label{eq:app_global_purity}
\end{equation}
Moreover, the Schmidt decomposition of a pure bipartite state ensures that $\hat{\rho}_{WE}$ and $\hat{\rho}_{CE}$ have the same nonzero eigenvalues and therefore
\begin{equation}
S(\hat{\rho}_{WE})=S(\hat{\rho}_{CE}).
\label{eq:app_equal_entropies}
\end{equation}
The quantum mutual information consequently becomes
\begin{equation}
I(WE:CE)_\rho
=S(\hat{\rho}_{WE})+S(\hat{\rho}_{CE})-S(\hat{\rho}_{WE,CE})
=2S(\hat{\rho}_{CE}).
\label{eq:app_mutual_information}
\end{equation}

To evaluate the classically accessible correlation for measurements on the WE, let $\{\Pi_k^{WE}\}$ be a complete set of rank-one orthogonal projectors. Acting on the global state gives
\begin{equation}
(\Pi_k^{WE}\otimes\mathbb{I}_{CE})\ket{\Psi}
=\ket{w_k}\otimes\ket{\phi_k},
\label{eq:app_projected_state}
\end{equation}
where $\ket{\phi_k}$ is an unnormalized CE state and $P_k=\braket{\phi_k|\phi_k}$ is the probability of outcome $k$. The conditional CE state is therefore
\begin{equation}
\hat{\rho}_{CE|k}
=\frac{\ket{\phi_k}\bra{\phi_k}}{\braket{\phi_k|\phi_k}},
\label{eq:app_conditional_state}
\end{equation}
which is pure for every outcome with $P_k>0$. Hence,
\begin{equation}
\sum_k P_k S(\hat{\rho}_{CE|k})=0,
\qquad
J(CE:WE)_\rho=S(\hat{\rho}_{CE}).
\label{eq:app_classical_correlation}
\end{equation}
Combining Eqs.~(\ref{eq:app_mutual_information}) and~(\ref{eq:app_classical_correlation}) gives
\begin{equation}
\mathcal{D}(CE:WE)_\rho
=I(WE:CE)_\rho-J(CE:WE)_\rho
=S(\hat{\rho}_{CE})\geq 0.
\label{eq:app_discord}
\end{equation}

At $p=1$, the state is exactly $\ket{S}$, and therefore $I=J=\mathcal{D}=0$. For $0<p<1$, a generic coefficient matrix $a_{ij}$ of Schmidt rank greater than one produces a nonproduct state, for which $S(\hat{\rho}_{CE})>0$ and all three correlation measures are positive. The special coefficient choices may produce isolated factorizable values of $p$. At $p=0$, the state reduces to $\ket{\mathcal{C}}$ and remains correlated only when $a_{ij}$ has a Schmidt rank greater than one. Thus, within the present pure-state construction, nonzero discord accompanies every nonproduct WE--CE state.

\bibliography{refs}

@misc{Miranda2024a,
  doi = {10.48550/ARXIV.2410.11928},
  url = {https://arxiv.org/abs/2410.11928},
  author = {Miranda, David A. and Pinz{\'o}n, Edgar F. and Bueno, Paulo R.},
  title = {Quantum Electrodynamics in an Electrolyte Medium Driving Entanglement Between Graphene Sheets},
  publisher = {arXiv},
  year = {2024},
  note = {Preprint}
}

@book{Bard2022,
  title = {Electrochemical Methods: Fundamentals and Applications},
  author = {Bard, Allen J. and Faulkner, Larry R. and White, Henry S.},
  year = {2022},
  edition = {3rd},
  publisher = {John Wiley \& Sons},
  address = {Hoboken, NJ},
  isbn = {978-1119334057}
}

@article{Snizhko2023Potentiostat,
  title = {Potentiostat Design Keys for Analytical Applications},
  author = {Snizhko, D. and Zholudov, Y. and Kukoba, A. and Xu, G.},
  journal = {Journal of Electroanalytical Chemistry},
  year = {2023},
  doi = {10.1016/j.jelechem.2023.117380}
}

@article{Vanysek2013The,
  title = {The Role of the Reference and Counter Electrodes in Electrochemical Impedance Measurement},
  author = {Van{\'y}sek, Petr and Tavassol, Hadi and Pilson, Kate-Leigh},
  journal = {ECS Meeting Abstracts},
  volume = {MA2013-02},
  number = {48},
  pages = {2684--2684},
  year = {2013},
  doi = {10.1149/ma2013-02/48/2684}
}

@article{Wang2021Electrochemical,
  title = {Electrochemical Impedance Spectroscopy},
  author = {Wang, Shangshang and Zhang, Jianbo and Gharbi, O. and Vivier, V. and Gao, Ming and Orazem, M.},
  journal = {Nature Reviews Methods Primers},
  year = {2021},
  volume = {1},
  doi = {10.1038/s43586-021-00039-w}
}

@article{Lazanas2023Electrochemical,
  title = {Electrochemical Impedance Spectroscopy: A Tutorial},
  author = {Lazanas, A. and Prodromidis, M.},
  journal = {ACS Measurement Science Au},
  year = {2023},
  volume = {3},
  pages = {162--193},
  doi = {10.1021/acsmeasuresciau.2c00070}
}

@article{Magar2021Electrochemical,
  title = {Electrochemical Impedance Spectroscopy (EIS): Principles, Construction, and Biosensing Applications},
  author = {Magar, Hend S. and Hassan, Rabeay Y. A. and Mulchandani, A.},
  journal = {Sensors},
  year = {2021},
  volume = {21},
  doi = {10.3390/s21196578}
}

@article{Huang2016Graphical,
  title = {Graphical Analysis of Electrochemical Impedance Spectroscopy Data in Bode and Nyquist Representations},
  author = {Huang, Jun and Li, Zhe and Liaw, B. and Zhang, Jianbo},
  journal = {Journal of Power Sources},
  year = {2016},
  volume = {309},
  pages = {82--98},
  doi = {10.1016/j.jpowsour.2016.01.073}
}

@article{Miranda2021Nanostructured,
  title = {Nanostructured Titanium Dioxide Surfaces for Electrochemical Biosensing},
  author = {Bertel, L. and Miranda, D. and Garc{\'i}a-Mart{\'i}n, J.},
  journal = {Sensors},
  year = {2021},
  volume = {21},
  doi = {10.3390/s21186167}
}

@article{Shen2018Titanium,
  title = {Titanium Dioxide Nanostructures for Photoelectrochemical Applications},
  author = {Shen, Shaohua and Chen, Jie and Wang, Meng and Sheng, Xia and Chen, Xiangyan and Feng, Xinjian and Mao, S.},
  journal = {Progress in Materials Science},
  year = {2018},
  doi = {10.1016/j.pmatsci.2018.07.006}
}

@article{Berger2012The,
  title = {The Electrochemistry of Nanostructured Titanium Dioxide Electrodes},
  author = {Berger, T. and Monllor-Satoca, D. and Jankulovska, M. and Lana-Villarreal, T. and G{\'o}mez, R.},
  journal = {ChemPhysChem},
  year = {2012},
  volume = {13},
  pages = {2824--2875},
  doi = {10.1002/cphc.201200073}
}

@article{Wu2022Understanding,
  author = {Wu, Jianzhong},
  title = {Understanding the Electric Double-Layer Structure, Capacitance, and Charging Dynamics},
  journal = {Chemical Reviews},
  volume = {122},
  number = {12},
  pages = {10821--10859},
  year = {2022},
  doi = {10.1021/acs.chemrev.2c00097}
}

@article{Gadipelli2023Understanding,
  author = {Gadipelli, Srinivas and Guo, Jian and Li, Zhuangnan and Howard, Christopher A. and Liang, Yini and Zhang, Hong and Shearing, Paul R. and Brett, Dan J. L.},
  title = {Understanding and Optimizing Capacitance Performance in Reduced Graphene-Oxide Based Supercapacitors},
  journal = {Small Methods},
  volume = {7},
  number = {6},
  pages = {e2201557},
  year = {2023},
  doi = {10.1002/smtd.202201557}
}

@article{Luryi1988QuantumCapacitance,
  author = {Luryi, Serge},
  title = {Quantum Capacitance Device},
  journal = {Applied Physics Letters},
  volume = {52},
  pages = {501--503},
  year = {1988},
  doi = {10.1063/1.99649}
}

@book{Bisquert2015,
  author = {Bisquert, J.},
  title = {Nanostructured Energy Devices},
  publisher = {CRC Press},
  year = {2015}
}

@article{RevModPhys.84.1655,
  title = {The Classical-Quantum Boundary for Correlations: Discord and Related Measures},
  author = {Modi, Kavan and Brodutch, Aharon and Cable, Hugo and Paterek, Tomasz and Vedral, Vlatko},
  journal = {Reviews of Modern Physics},
  volume = {84},
  pages = {1655--1707},
  year = {2012},
  doi = {10.1103/RevModPhys.84.1655}
}

@article{Adesso2016Measures,
  title = {Measures and Applications of Quantum Correlations},
  author = {Adesso, G. and Bromley, T. and Cianciaruso, M.},
  journal = {Journal of Physics A: Mathematical and Theoretical},
  year = {2016},
  volume = {49},
  doi = {10.1088/1751-8113/49/47/473001}
}

@article{Lesne2014Shannon,
title={Shannon entropy: a rigorous notion at the crossroads between probability, information theory, dynamical systems and statistical physics},
author={A. Lesne},
journal={Mathematical Structures in Computer Science},
year={2014},
volume={24},
doi={10.1017/s0960129512000783}
}

@book{blum2012density,
  title={Density Matrix Theory and Applications},
  author={Blum, Karl},
  edition={3rd},
  year={2012},
  publisher={Springer},
  series={Springer Series on Atomic, Optical, and Plasma Physics},
  volume={64},
  isbn={978-3-642-20560-6}
}

@book{nielsen_chuang_2010,
  title={Quantum Computation and Quantum Information: 10th Anniversary Edition},
  author={Nielsen, Michael A. and Chuang, Isaac L.},
  year={2010},
  publisher={Cambridge University Press},
  address={Cambridge},
  isbn={978-1-107-00217-3}
}

@article{Ollivier2001_Quantum,
  title = {Quantum Discord: A Measure of the Quantumness of Correlations},
  author = {Ollivier, Harold and Zurek, Wojciech H.},
  journal = {Phys. Rev. Lett.},
  volume = {88},
  issue = {1},
  pages = {017901},
  numpages = {4},
  year = {2001},
  month = {Dec},
  publisher = {American Physical Society},
  doi = {10.1103/PhysRevLett.88.017901},
  url = {https://link.aps.org/doi/10.1103/PhysRevLett.88.017901}
}

@article{Zhang2023,
  title = {Reliable impedance analysis of Li-ion battery half-cell by standardization on electrochemical impedance spectroscopy (EIS)},
  volume = {158},
  ISSN = {1089-7690},
  url = {http://dx.doi.org/10.1063/5.0139347},
  DOI = {10.1063/5.0139347},
  number = {5},
  journal = {The Journal of Chemical Physics},
  publisher = {AIP Publishing},
  author = {Zhang,  Baodan and Wang,  Lingling and Zhang,  Yiming and Wang,  Xiaotong and Qiao,  Yu and Sun,  Shi-Gang},
  year = {2023},
  month = Feb 
}

@article{Li2022,
  title = {Impedance response of electrochemical interfaces. III. Fingerprints of couplings between interfacial electron transfer reaction and electrolyte-phase ion transport},
  volume = {157},
  ISSN = {1089-7690},
  url = {http://dx.doi.org/10.1063/5.0119592},
  DOI = {10.1063/5.0119592},
  number = {18},
  journal = {The Journal of Chemical Physics},
  publisher = {AIP Publishing},
  author = {Li,  Chen Kun and Zhang,  Jianbo and Huang,  Jun},
  year = {2022},
  month = Nov 
}

@article{Yamamoto2024,
  title = {Polarization of disk electrodes in high-conductivity electrolyte solutions},
  volume = {160},
  ISSN = {1089-7690},
  url = {http://dx.doi.org/10.1063/5.0179083},
  DOI = {10.1063/5.0179083},
  number = {5},
  journal = {The Journal of Chemical Physics},
  publisher = {AIP Publishing},
  author = {Yamamoto,  Kenneth K. and Koklu,  Anil and Beskok,  Ali and Ajaev,  Vladimir S.},
  year = {2024},
  month = Feb 
}

@article{DiPasquale2023,
  title = {Constant chemical potential–quantum mechanical–molecular dynamics simulations of the graphene–electrolyte double layer},
  volume = {158},
  ISSN = {1089-7690},
  url = {http://dx.doi.org/10.1063/5.0138267},
  DOI = {10.1063/5.0138267},
  number = {13},
  journal = {The Journal of Chemical Physics},
  publisher = {AIP Publishing},
  author = {Di Pasquale,  Nicodemo and Finney,  Aaron R. and Elliott,  Joshua D. and Carbone,  Paola and Salvalaglio,  Matteo},
  year = {2023},
  month = Apr 
}

@article{Li2023,
  title = {Molecular understanding of the Helmholtz capacitance difference between Cu(100) and graphene electrodes},
  volume = {158},
  ISSN = {1089-7690},
  url = {http://dx.doi.org/10.1063/5.0139534},
  DOI = {10.1063/5.0139534},
  number = {8},
  journal = {The Journal of Chemical Physics},
  publisher = {AIP Publishing},
  author = {Li,  Xiang-Ying and Jin,  Xiang-Feng and Yang,  Xiao-Hui and Wang,  Xue and Le,  Jia-Bo and Cheng,  Jun},
  year = {2023},
  month = Feb 
}

@article{Coquinot2026,
  title = {Electron–electrolyte coupling in AC transport through nanofluidic channels},
  volume = {164},
  ISSN = {1089-7690},
  url = {http://dx.doi.org/10.1063/5.0313352},
  DOI = {10.1063/5.0313352},
  number = {13},
  journal = {The Journal of Chemical Physics},
  publisher = {AIP Publishing},
  author = {Coquinot,  Baptiste and Lizée,  Mathieu and Bocquet,  Lydéric and Kavokine,  Nikita},
  year = {2026},
  month = Apr 
}

@article{Chakraborty2024Rectification,
title={Rectification of high-frequency artifacts in EIS data of three-electrode Li-ion cells},
author={Arup Chakraborty and Tazdin Amietszajew},
journal={Electrochimica Acta},
year={2024},
doi={10.1016/j.electacta.2024.145266}}

@article{Levi2014Impedance,title={Impedance Spectra of Energy-Storage Electrodes Obtained with Commercial Three-Electrode Cells: Some Sources of Measurement Artefacts},author={M. Levi and Vadim Dargel and Y. Shilina and D. Aurbach and I. Halalay},journal={Electrochimica Acta},year={2014},volume={149},pages={126-135},doi={10.1016/j.electacta.2014.10.083}}

@article{Tatara2018The,title={The Effect of Electrode-Electrolyte Interface on the Electrochemical Impedance Spectra for Positive Electrode in Li-Ion Battery},author={R. Tatara and Pinar Karayaylali and Yang Yu and Y. Zhang and L. Giordano and F. Maglia and Roland Jung and J. Schmidt and Isaac Lund and Y. Shao-horn},journal={Journal of The Electrochemical Society},year={2018},doi={10.1149/2.0121903jes}}

@article{Wang2019Nonequilibrium,title={Nonequilibrium effects on quantum correlations: Discord, mutual information, and entanglement of a two-fermionic system in bosonic and fermionic environments},author={Xuanhua Wang and Jin Wang},journal={Physical Review A},year={2019},doi={10.1103/physreva.100.052331}}

@article{Luo2008Quantum,title={Quantum discord for two-qubit systems},author={S. Luo},journal={Physical Review A},year={2008},volume={77},pages={042303},doi={10.1103/physreva.77.042303}}

@article{Miao2023Entanglement,title={Entanglement and quantum discord in the cavity QED models},author={Huimin Miao and Wanshun Li},journal={Heliyon},year={2023},volume={11},doi={10.1016/j.heliyon.2024.e41194}}

@article{Jadoon2024Hybrid,title={Hybrid TiO2–RGO nanocomposite as high specific capacitance electrode for supercapacitor},author={Jamil K Jadoon and Phuong V. Pham},journal={Nanotechnology},year={2024},volume={35},doi={10.1088/1361-6528/ad6a6a}}

@article{Tayebi2019Reduced,title={Reduced graphene oxide (RGO) on TiO2 for an improved photoelectrochemical (PEC) and photocatalytic activity},author={Meysam Tayebi and M. Kolaei and A. Tayyebi and Z. Masoumi and Z. Belbasi and Byeong-Kyu Lee},journal={Solar Energy},year={2019},doi={10.1016/j.solener.2019.08.020}}

@article{Hahn2019,
  title = {Optimized Process Parameters for a Reproducible Distribution of Relaxation Times Analysis of Electrochemical Systems},
  volume = {5},
  ISSN = {2313-0105},
  url = {http://dx.doi.org/10.3390/batteries5020043},
  DOI = {10.3390/batteries5020043},
  number = {2},
  journal = {Batteries},
  publisher = {MDPI AG},
  author = {Hahn,  Markus and Schindler,  Stefan and Triebs,  Lisa-Charlotte and Danzer,  Michael A.},
  year = {2019},
  month = May,
  pages = {43}
}

@article{ITAGAKI2007,
  title = {Electrochemical Impedance and Complex Capacitance to Interpret Electrochemical Capacitor},
  volume = {75},
  ISSN = {1344-3542},
  url = {http://dx.doi.org/10.5796/electrochemistry.75.649},
  DOI = {10.5796/electrochemistry.75.649},
  number = {8},
  journal = {Electrochemistry},
  publisher = {The Electrochemical Society of Japan},
  author = {ITAGAKI,  Masayuki and SUZUKI,  Satoshi and SHITANDA,  Isao and WATANABE,  Kunihiro},
  year = {2007},
  pages = {649–655}
}

@misc{GTM2026a,
  author    = {Gonzalez, Kevin A. and Toledo, Nicol{\'a}s H. and Miranda, David A.},
  title     = {Data and code for: Counter-Electrode Dependence of the Working-Electrode Capacitance in Three-Electrode {TiO2 + rGO} Cells: Classical Controls and Interpretation within a Quantum-Discord Framework},
  year      = {2026},
  month     = aug,
  publisher = {Zenodo},
  version   = {1.0.0},
  doi       = {10.5281/zenodo.22178264},
  url       = {https://doi.org/10.5281/zenodo.22178264},
  copyright = {Creative Commons Attribution 4.0 International}
}
\end{document}

\title{Supplementary Material for ``Counter-Electrode Dependence of the Working-Electrode Capacitance in Three-Electrode $TiO_2+rGO$ Cells: Classical Controls and Interpretation within a Quantum-Discord Framework''}

\author{Kevin A. Gonzalez}
\author{Nicolás H. Toledo}
\author{David A. Miranda}
\email{dalemir@uis.edu.co}
\affiliation{Universidad Industrial de Santander, 680002 Bucaramanga, Santander, Colombia}
\date{\today}
\maketitle

\section*{Contents}

This Supplementary Material contains the full information-theoretic derivations used to define the classical two-electrode and three-electrode reference cases in the main manuscript, together with expanded algebra supporting the formal pure-state derivation presented in Appendix A.

\section{Classical Two-Electrode Reference}

The two-electrode configuration is the classical reference in which the counter-electrode contribution is intentionally included in the measured observable. The measured impedance is
\begin{equation}
\label{eq:si_two_electrode}
Z_m^{(2p)}(\omega)=Z_{WE}(\omega)+Z_{CE}(\omega).
\end{equation}

Let the relevant random variables be the measured two-electrode impedance, $Z_m^{(2p)}$, and the independently characterized counter-electrode impedance, $Z_{CE}$. Before imposing the two-electrode measurement constraint, the local working-electrode and counter-electrode impedance contributions may be taken as statistically independent,
\begin{equation}
\label{eq:si_independent_local_impedances}
p\!\left(z_{WE},z_{CE}\right)
=
p(z_{WE})p\!\left(z_{CE}\right).
\end{equation}
This independence does not imply that the derived measured variable $Z_m^{(2p)}$ is independent of $Z_{CE}$, because $Z_m^{(2p)}$ contains the counter-electrode contribution.

The mutual information between the measured two-electrode impedance and the independently characterized counter-electrode impedance is
\begin{equation}
\label{eq:si_mi_two_electrode}
I\!\left(Z_m^{(2p)};Z_{CE}\right)
=
\sum\limits_{z_m^{(2p)},z_{CE}}
p\!\left(z_m^{(2p)},z_{CE}\right)
\ln
\left[
\frac{
p\!\left(z_m^{(2p)},z_{CE}\right)
}{
p\!\left(z_m^{(2p)}\right)
p\!\left(z_{CE}\right)
}
\right].
\end{equation}
The non-negativity of mutual information and its vanishing for statistically independent variables follow from standard information-theoretic properties of Shannon entropy \cite{Lesne2014Shannon}.

Under the additive classical constraint, $z_m^{(2p)}=z_{WE}+z_{CE}$. For discrete variables, the corresponding joint distribution can be written as
\begin{equation}
\label{eq:si_additive_joint_constraint}
p\!\left(z_m^{(2p)},z_{CE}\right)
=
\sum\limits_{z_{WE}}
p(z_{WE})
p\!\left(z_{CE}\right)
\delta\!\left(
z_m^{(2p)}
-
z_{WE}
-
z_{CE}
\right).
\end{equation}
Equivalently, after imposing the additive constraint,
\begin{equation}
\label{eq:si_additive_condensed}
p\!\left(z_m^{(2p)},z_{CE}\right)
=
p\!\left(z_{CE}\right)
p\!\left(z_m^{(2p)}-z_{CE}\right),
\end{equation}
where $p\!\left(z_m^{(2p)}-z_{CE}\right)$ represents the probability of the working-electrode contribution compatible with the measured total impedance.

The classically accessible correlation can be written as
\begin{equation}
\label{eq:si_two_electrode_j}
J\!\left(Z_m^{(2p)}|Z_{CE}\right)
=
H\!\left(Z_m^{(2p)}\right)
-
H\!\left(Z_m^{(2p)}|Z_{CE}\right).
\end{equation}
Using the definition of conditional entropy,
\begin{equation*}
H\!\left(Z_m^{(2p)}|Z_{CE}\right)
=
-
\sum\limits_{z_m^{(2p)},z_{CE}}
p\!\left(z_m^{(2p)},z_{CE}\right)
\ln p\!\left(z_m^{(2p)}|z_{CE}\right),
\end{equation*}
and
\begin{equation*}
H\!\left(Z_m^{(2p)}\right)
=
-
\sum\limits_{z_m^{(2p)}}
p\!\left(z_m^{(2p)}\right)
\ln p\!\left(z_m^{(2p)}\right),
\end{equation*}
one obtains
\begin{equation}
\label{eq:si_two_electrode_j_equals_i}
\begin{aligned}
J\!\left(Z_m^{(2p)}|Z_{CE}\right)
&=
\sum\limits_{z_m^{(2p)},z_{CE}}
p\!\left(z_m^{(2p)},z_{CE}\right)
\ln
\left[
\frac{
p\!\left(z_m^{(2p)}|z_{CE}\right)
}{
p\!\left(z_m^{(2p)}\right)
}
\right] \\
&=
\sum\limits_{z_m^{(2p)},z_{CE}}
p\!\left(z_m^{(2p)},z_{CE}\right)
\ln
\left[
\frac{
p\!\left(z_m^{(2p)},z_{CE}\right)
}{
p\!\left(z_m^{(2p)}\right)
p\!\left(z_{CE}\right)
}
\right] \\
&=
I\!\left(Z_m^{(2p)};Z_{CE}\right).
\end{aligned}
\end{equation}
Thus, the classical two-electrode measurement defines a positive classical-correlation reference case whenever CE variation is experimentally distinguishable. A nonzero mutual information in this configuration does not indicate a nonclassical effect; it reflects the fact that the CE contribution is part of the measured two-electrode observable.

\section{Classical Three-Electrode Reference}

The three-electrode configuration defines the classical reference case in which the CE closes the current path but its impedance is not part of the measured WE--RE observable. Under ideal potentiostatic operation,
\begin{equation}
\label{eq:si_three_electrode}
Z_m^{(3p)}(\omega)=Z_{WE}(\omega).
\end{equation}

Let repeated EIS measurements be represented by the measured three-electrode impedance, $Z_m^{(3p)}$, and the independently characterized counter-electrode impedance, $Z_{CE}$. The classical three-electrode expectation states that changes in $Z_{CE}$ should not modify the measured WE response once ordinary electrochemical and instrumental artifacts are excluded. Therefore,
\begin{equation}
\label{eq:si_three_electrode_factorization}
p\!\left(z_m^{(3p)},z_{CE}\right)
=
p\!\left(z_m^{(3p)}\right)
p\!\left(z_{CE}\right).
\end{equation}

The mutual information between the measured three-electrode response and the independently characterized CE state is
\begin{equation}
\label{eq:si_mi_three_electrode}
I\!\left(Z_m^{(3p)};Z_{CE}\right)
=
\sum\limits_{z_m^{(3p)},z_{CE}}
p\!\left(z_m^{(3p)},z_{CE}\right)
\ln
\left[
\frac{
p\!\left(z_m^{(3p)},z_{CE}\right)
}{
p\!\left(z_m^{(3p)}\right)
p\!\left(z_{CE}\right)
}
\right].
\end{equation}
Using the factorization in Eq.~(\ref{eq:si_three_electrode_factorization}), the logarithmic term becomes $\ln(1)=0$, so
\begin{equation}
\label{eq:si_three_electrode_mi_zero}
I\!\left(Z_m^{(3p)};Z_{CE}\right)=0.
\end{equation}

The same conclusion follows for the classically accessible correlation:
\begin{equation}
\label{eq:si_three_electrode_j}
J\!\left(Z_m^{(3p)}|Z_{CE}\right)
=
H\!\left(Z_m^{(3p)}\right)
-
H\!\left(Z_m^{(3p)}|Z_{CE}\right).
\end{equation}
If $Z_m^{(3p)}$ and $Z_{CE}$ are statistically independent, knowledge of the CE state does not reduce the uncertainty of the measured WE--RE response, so
\begin{equation}
\label{eq:si_three_electrode_zero_correlation}
J\!\left(Z_m^{(3p)}|Z_{CE}\right)
=
I\!\left(Z_m^{(3p)};Z_{CE}\right)
=0.
\end{equation}
This zero-correlation result does not mean that the CE is physically inactive. The CE supplies the current required by the potentiostatic control loop. Rather, it means that the independently characterized CE impedance should not be encoded in the measured WE--RE impedance under the ideal classical three-electrode measurement model.

The experimentally testable expectation is
\begin{equation}
\label{eq:si_delta_ce_reference}
\Delta Z_{CE}(\omega)
=
Z_m^{(3p)}(\omega|CE=TiO_2+rGO)
-
Z_m^{(3p)}(\omega|CE=Pt)
=0,
\end{equation}
within uncertainty and after ordinary classical artifacts have been excluded. A systematic nonzero $\Delta Z_{CE}(\omega)$, or a nonzero $I\!\left(Z_m^{(3p)};Z_{CE}\right)$, identifies a deviation from the ideal classical three-electrode expectation. It does not by itself prove quantum discord or a microscopic quantum state; it identifies a macroscopic impedance signature requiring interpretation beyond the simple independent-impedance model.

\section{Expanded Algebra Supporting Appendix A}

This section provides the component-level algebra underlying the normalization, reduced-state, and conditional-measurement relations presented in Appendix A of the main manuscript. The conceptual derivation and the resulting expressions for quantum mutual information, classically accessible correlation, and quantum discord are retained in the Appendix.

\subsection{Coefficient Representation and Normalization}

Let the normalized product and correlated contributions be
\begin{equation}
\ket{S}=\ket{WE}\ket{CE}
=\sum_{i,j}w_i c_j\ket{WE_i}\ket{CE_j},
\qquad
\ket{\mathcal{C}}=\sum_{i,j}a_{ij}\ket{WE_i}\ket{CE_j},
\label{eq:sm_components}
\end{equation}
where $\sum_i|w_i|^2=\sum_j|c_j|^2=\sum_{i,j}|a_{ij}|^2=1$. Their overlap is
\begin{equation}
z\equiv\braket{S|\mathcal{C}}
=\sum_{i,j}w_i^*c_j^*a_{ij},
\qquad
\braket{\mathcal{C}|S}=z^*.
\label{eq:sm_overlap}
\end{equation}
For the ansatz $\ket{\Psi}=\sqrt{p}\ket{S}+\sqrt{1-p}\ket{\mathcal{C}}$, direct expansion gives
\begin{align}
\braket{\Psi|\Psi}
&=p\braket{S|S}
+\sqrt{p(1-p)}\braket{S|\mathcal{C}}
+\sqrt{p(1-p)}\braket{\mathcal{C}|S}
+(1-p)\braket{\mathcal{C}|\mathcal{C}} \notag\\
&=1+2\sqrt{p(1-p)}\,\operatorname{Re}(z).
\label{eq:sm_norm_expansion}
\end{align}
Thus, $\operatorname{Re}(z)=0$ ensures $\braket{\Psi|\Psi}=1$ for every $p$.

Introducing
\begin{equation}
B_{mn}=\sqrt{p}\,w_m c_n+\sqrt{1-p}\,a_{mn},
\label{eq:sm_Bmn}
\end{equation}
the normalized state becomes
\begin{equation}
\ket{\Psi}=\sum_{m,n}B_{mn}\ket{WE_m}\ket{CE_n},
\qquad
\sum_{m,n}|B_{mn}|^2=1.
\label{eq:sm_state_Bmn}
\end{equation}

\subsection{Explicit Partial Traces}

The composite density operator is
\begin{equation}
\hat{\rho}_{WE,CE}
=
\sum_{m,n,m',n'}
B_{mn}B_{m'n'}^*
\left(
\ket{WE_m}\bra{WE_{m'}}
\otimes
\ket{CE_n}\bra{CE_{n'}}
\right).
\label{eq:sm_density_expansion}
\end{equation}
Using $\operatorname{Tr}_{CE}(\ket{CE_n}\bra{CE_{n'}})=\delta_{nn'}$ gives
\begin{align}
\hat{\rho}_{WE}
&=\operatorname{Tr}_{CE}(\hat{\rho}_{WE,CE}) \notag\\
&=\sum_{m,m'}
\left(
\sum_n B_{mn}B_{m'n}^*
\right)
\ket{WE_m}\bra{WE_{m'}}.
\label{eq:sm_rho_we}
\end{align}
Likewise, $\operatorname{Tr}_{WE}(\ket{WE_m}\bra{WE_{m'}})=\delta_{mm'}$ yields
\begin{align}
\hat{\rho}_{CE}
&=\operatorname{Tr}_{WE}(\hat{\rho}_{WE,CE}) \notag\\
&=\sum_{n,n'}
\left(
\sum_m B_{mn}B_{mn'}^*
\right)
\ket{CE_n}\bra{CE_{n'}}.
\label{eq:sm_rho_ce}
\end{align}
These component expressions are the reduced density operators used in Appendix A.

\subsection{Conditional CE State after a WE Measurement}

Let $\Pi_k^{WE}=\ket{w_k}\bra{w_k}$ be a rank-one projector on the WE subsystem. Acting on the bipartite state gives
\begin{align}
(\Pi_k^{WE}\otimes\mathbb{I}_{CE})\ket{\Psi}
&=
\sum_{i,j}B_{ij}
\ket{w_k}\braket{w_k|WE_i}\ket{CE_j} \notag\\
&=\ket{w_k}\otimes\ket{\phi_k},
\label{eq:sm_projected_state}
\end{align}
where
\begin{equation}
\ket{\phi_k}
=
\sum_{i,j}B_{ij}\braket{w_k|WE_i}\ket{CE_j}.
\label{eq:sm_phi_k}
\end{equation}
The probability of outcome $k$ is
\begin{equation}
P_k
=
\operatorname{Tr}\!\left[
(\Pi_k^{WE}\otimes\mathbb{I}_{CE})\hat{\rho}_{WE,CE}
\right]
=
\braket{\phi_k|\phi_k}.
\label{eq:sm_probability}
\end{equation}
For $P_k>0$, tracing out the WE gives
\begin{equation}
\hat{\rho}_{CE|k}
=
\frac{
\operatorname{Tr}_{WE}\!\left[
(\Pi_k^{WE}\otimes\mathbb{I}_{CE})
\hat{\rho}_{WE,CE}
(\Pi_k^{WE}\otimes\mathbb{I}_{CE})
\right]
}{P_k}
=
\frac{\ket{\phi_k}\bra{\phi_k}}
{\braket{\phi_k|\phi_k}}.
\label{eq:sm_conditional_state}
\end{equation}
The conditional state is therefore rank one and has zero von Neumann entropy. This explicit calculation supports the result $J(CE:WE)_\rho=S(\hat{\rho}_{CE})$ derived in Appendix A.

\raggedbottom
\section{Additional Three-Electrode Counter-Electrode-Dependence Measurements}
\label{sec:additional_ce_dependence}

This section reports three-electrode measurements for the remaining four valid $TiO_2+rGO$ bar pairs, which complement the representative pair in the main text. No dummy element was used in these measurements. For each pair, Configuration 1 (Pt CE) is compared with Configuration 3 ($TiO_2+rGO$ CE). In Figs.~S1--S4, panel (a) shows the complex-capacitance Nyquist plot ($C''$ vs $C'$), and panel (b) the Bode plots ($C'$ top, $C''$ bottom) vs frequency. Configuration 1 spectra exhibit pair-dependent relaxations: several show a low-frequency dispersion near $10~\mathrm{Hz}$ and a higher-frequency contribution extending toward $\sim1~\mathrm{kHz}$, with varying positions and prominence. Configuration 3 spectra also vary in magnitude and structure. However, all four pairs show a qualitative CE-dependent difference consistent with that observed for the main-text pair, indicating that the CE dependence is not limited to a single bar pair.

\begin{figure}[t]
\centering
\includegraphics[width=\textwidth]{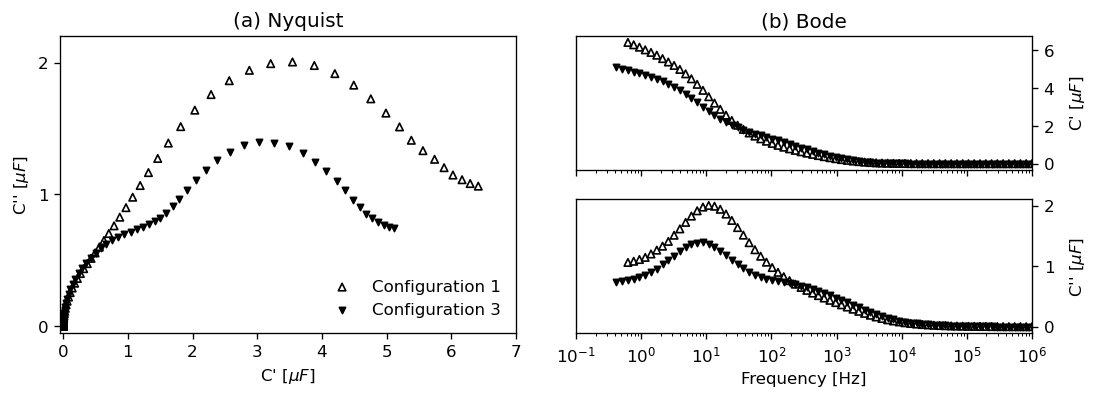}
\caption{Additional three-electrode counter-electrode-dependence measurement for a second $TiO_2+rGO$ bar pair. The Configuration 1 response with Pt as the CE is compared with the Configuration 3 response with a second $TiO_2+rGO$ bar as the CE. (a) Complex-capacitance Nyquist representation, $C''$ versus $C'$. (b) Bode representation showing $C'$ in the upper panel and $C''$ in the lower panel as functions of frequency. The two configurations exhibit a qualitative CE-dependent difference consistent with that observed for the representative pair in the main text.}
\label{fig:SI_additional_CE_dependence_1}
\end{figure}

\begin{figure}[t]
\centering
\includegraphics[width=\textwidth]{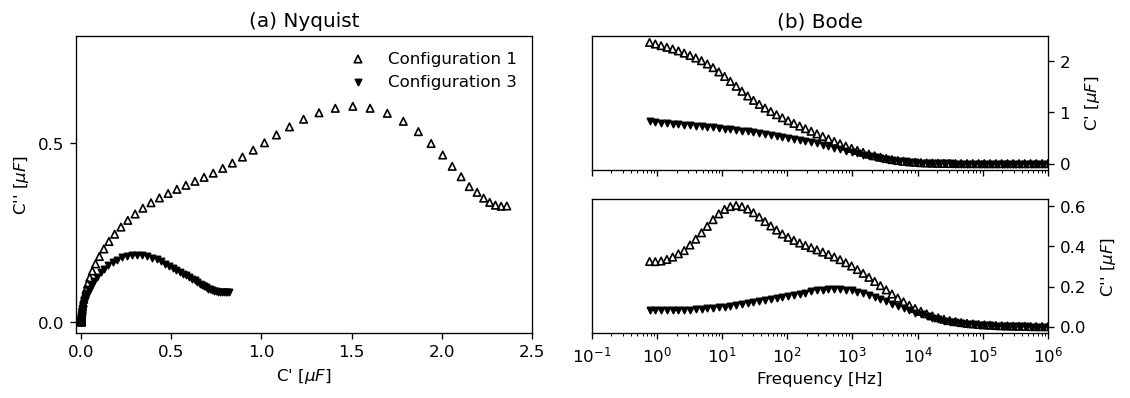}
\caption{Additional three-electrode counter-electrode-dependence measurement for a third $TiO_2+rGO$ bar pair. The Configuration 1 response with Pt as the CE is compared with the Configuration 3 response with a second $TiO_2+rGO$ bar as the CE. (a) Complex-capacitance Nyquist representation, $C''$ versus $C'$. (b) Bode representation showing $C'$ in the upper panel and $C''$ in the lower panel as functions of frequency. A qualitative CE-dependent difference is observed, with relaxation features whose positions and relative prominence differ from those of the other pairs.}
\label{fig:SI_additional_CE_dependence_2}
\end{figure}

\begin{figure}[H]
\centering
\includegraphics[width=\textwidth]{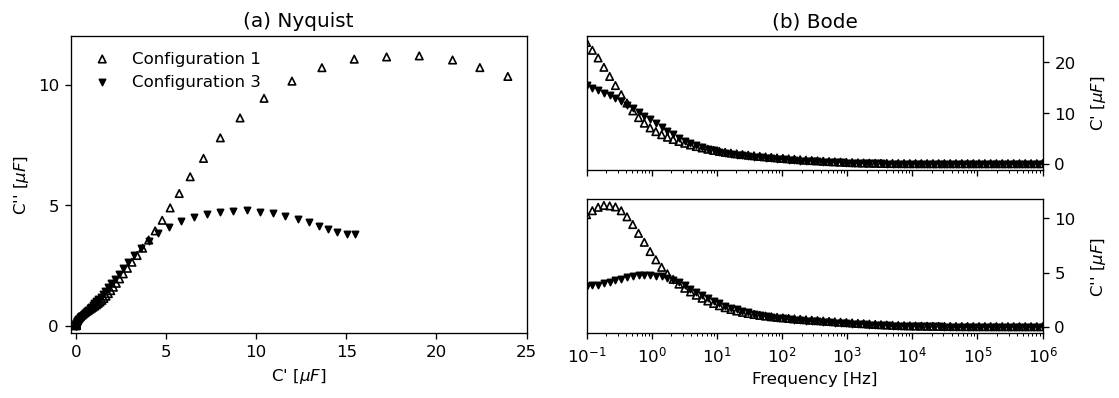}
\caption{Additional three-electrode counter-electrode-dependence measurement for a fourth $TiO_2+rGO$ bar pair. The Configuration 1 response with Pt as the CE is compared with the Configuration 3 response with a second $TiO_2+rGO$ bar as the CE. (a) Complex-capacitance Nyquist representation, $C''$ versus $C'$. (b) Bode representation showing $C'$ in the upper panel and $C''$ in the lower panel as functions of frequency. The observed difference between the two configurations is qualitatively consistent with that reported in the main text, while the absolute magnitude and relaxation structure remain pair dependent.}
\label{fig:SI_additional_CE_dependence_3}
\end{figure}

\begin{figure}[t]
\centering
\includegraphics[width=\textwidth]{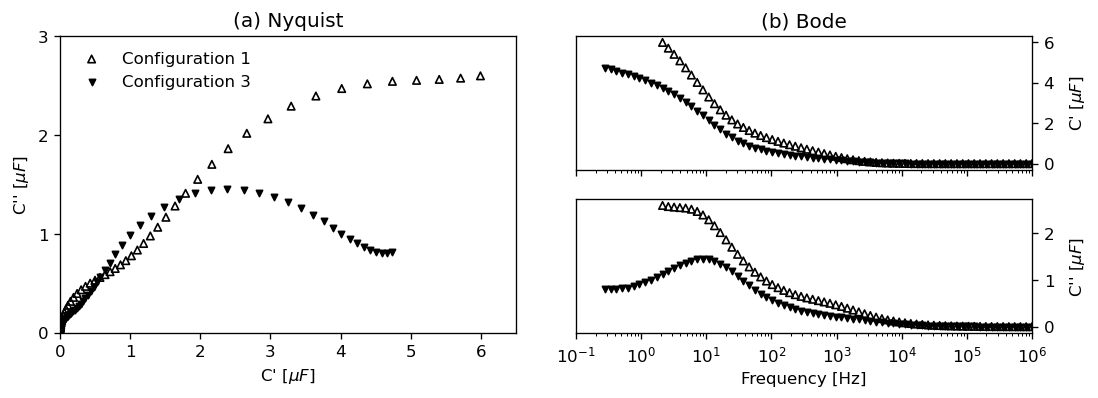}
\caption{Additional three-electrode counter-electrode-dependence measurement for a fifth $TiO_2+rGO$ bar pair. The Configuration 1 response with Pt as the CE is compared with the Configuration 3 response with a second $TiO_2+rGO$ bar as the CE. (a) Complex-capacitance Nyquist representation, $C''$ versus $C'$. (b) Bode representation showing $C'$ in the upper panel and $C''$ in the lower panel as functions of frequency. A qualitative CE-dependent difference is again observed, with pair-specific spectral magnitude and relaxation features.}
\label{fig:SI_additional_CE_dependence_4}
\end{figure}

\bibliography{refs}